%% file: 000_main.tex
\documentclass[sigconf]{aidb2021}
\newcommand\vldbdoi{XX.XX/XXX.XX}

\newcommand\vldbavailabilityurl{http://vldb.org/pvldb/format_vol14.html}
\newcommand\vldbpagestyle{plain} 

\usepackage{graphicx}
\usepackage{subcaption}
\usepackage{textcomp}
\usepackage{array}
\usepackage{multirow}
\usepackage{arydshln}
\usepackage{booktabs}
\usepackage{listings}
\usepackage{color}

\usepackage{microtype}
\usepackage{etoolbox}
\usepackage{refcount}
\usepackage{totpages}
\usepackage{environ}

\usepackage{xkeyval}
\usepackage{xstring}
\usepackage{setspace}
\usepackage{textcase}
\usepackage[bookmarksnumbered,unicode]{hyperref}
\usepackage{graphicx}
\usepackage[prologue]{xcolor}
\usepackage{geometry}
\usepackage{manyfoot}
\usepackage{iftex}
\usepackage{cmap}
\usepackage{balance}
\usepackage[square,numbers]{natbib}

\def \showarticletitle #1{#1}
\def \showISBNx     #1{\unskip}
\def \showISSN      #1{\unskip}
\newcommand\showeprint[2][arxiv]{%
  \def\@tempa{#1}%
  \ifx\@tempa\@empty\def\@tempa{arxiv}\fi
  \def\@tempb{arxiv}%
  \ifx\@tempa\@tempb
     arXiv:\href{https://arxiv.org/abs/#2}{#2}\else arXiv:#2%
  \fi}
\def \showURL       {\relax}
\newcommand{\newblock}{}

\begin{document}
\title{Machine Learning-based Selection of Graph Partitioning Strategy Using the Characteristics of Graph Data and Algorithm (Regular Papers)}

\author{
  YoungJoon Park\\
  \texttt{Department of Electrical \&} \\
  \texttt{Computer Engineering}\\
  \texttt{Seoul National University}\\
  \texttt{Seoul, South Korea}\\
  \texttt{dudwns930@snu.ac.kr}
  \and
  DongKyu Lee\\
  \texttt{Department of Electrical \&} \\
  \texttt{Computer Engineering}\\
  \texttt{Seoul National University}\\
  \texttt{Seoul, South Korea}\\
  \texttt{ardor12@snu.ac.kr}
  \and
  Tien-Cuong Bui\\
  \texttt{Department of Electrical \&} \\
  \texttt{Computer Engineering}\\
  \texttt{Seoul National University}\\
  \texttt{Seoul, South Korea}\\
  \texttt{cuongbt91@snu.ac.kr}
}






\maketitle
\input{00_abstract}

\pagestyle{\vldbpagestyle}
\begingroup\small\noindent\raggedright\textbf{AIDB Workshop Reference Format:}\\
YoungJoon Park, DongKyu Lee, Tien-Cuong Bui. Machine Learning-based Selection of Graph Partitioning Strategy Using the Characteristics of Graph Data and Algorithm. \textit{AIDB} 2021.
\endgroup
\boilerplate{}


\input{01_introduction} 
\input{01-05_Symbols}

\input{03_distributed_graph_engine}
\input{04_my_model}
\input{05_experiment}
\input{02_related_work}
\input{06_conclusion}
\input{07_acknowledgement}



\renewcommand\bibname{REFERENCES}
\renewcommand\refname{REFERENCES}
\newcounter{bibcount}
\makeatletter
\renewcommand\bibsection{%
  \section{\refname}%
}%
\renewcommand\@biblabel[1]{[#1]}
\makeatother

\end{document}

%% file: 00_abstract.tex
\abstract
Analyzing large graph data is an essential part of many modern applications, such as social networks. 
Due to its large computational complexity, distributed processing is frequently employed. 
This requires graph data to be divided across nodes, and the choice of partitioning strategy has a great impact on the execution time of the task. 
Yet, there is no one-size-fits-all partitioning strategy that performs well on arbitrary graph data and algorithms. 
The performance of a strategy depends on the characteristics of the graph data and algorithms. 
Moreover, due to the complexity of graph data and algorithms, manually identifying the best partitioning strategy is also infeasible. 
In this work, we propose a machine learning-based approach to select the most appropriate partitioning strategy 
for a given graph and processing algorithm. 
Our approach enumerates viable partitioning strategies, predicts the execution time of the target algorithm for each, and selects the partitioning strategy with the fastest estimated execution time. 
Our machine learning model is trained on features extracted from graph data and algorithm pseudo-code. 
We also propose a method that augments real execution logs of graph tasks to create a large synthetic dataset. 
Evaluation results show that the strategies selected by our approach lead to 1.46$\times$ faster execution time on average compared with the mean execution time of the partitioning strategies and about 0.95$\times$ the performance compared to the best partitioning strategy. 

%% file: 01_introduction.tex
\section{Introduction}
Graph data are prevalent in various fields, such as social networks~\cite{258876}, protein structures~\cite{10.5555/3294996.3295074}, web structures~\cite{10.1145/335168.335170}, textual structures~\cite{10.1145/2808797.2808872}, and e-commerce~\cite{10.1145/3219819.3219869}. 
As the amount of graph data increases fast, distributed computing of graph analysis can be an effective approach for large-scale graph data.
For example, it takes more than 10,000 seconds to calculate the local clustering coefficient of each vertex for the Clueweb12 data\cite{callan2012lemur} which has about 6.3 billion vertices and about 66.8 billion edges using 25 machines\cite{10.1145/3183713.3196915}.


There are several kinds of research about distributed graph processing. 
First, partitioning strategies \cite{10.1145/2806416.2806424, chen2019powerlyra, 10.1145/2484425.2484429, xie2014distributed, 10.1145/2556195.2556213, nazi2019gap} were proposed to partition graph data into a cluster. 
Second, distributed graph processing engines \cite{chen2019powerlyra, 10.5555/2387880.2387883, 10.1145/1807167.1807184, 10.5555/2685048.2685096, 10.1145/2484425.2484427, noauthororeditorneo4j} emerged to analyze distributed graph data. 
Finally, parallel algorithms \cite{10.1007/11735106_22, jones1993parallel, 10.1145/2187836.2187963, 10.1145/2925426.2926287} emerged to exploit the distributed environment. 
We focus on selecting the best partitioning strategy. 


A partitioning strategy determines how vertices and edges are divided into clusters, with the main differentiating points being communication cost, computation time, and replication factor which means the ratio of the number of the replicated vertex to the number of the original vertex. 
Existing partitioning strategies can be categorized into model agnostic, edge-cut partitioning, and vertex-cut partitioning\cite{abbas2018streaming}, where each may consider locality and/or load-balancing. In this paper, we define a \textit{\textbf{task}} to be a job that performs a specific algorithm on a specific graph, and the \textit{\textbf{performance}} of a partitioning strategy as the execution time of a task under the partitioning strategy after partitioning has finished.

\input{figure/Fig1_motivation}
The motivation for our research is that the performance of a partitioning strategy is different depending on a task.
Figure \ref{fig1:main} represents execution times of some tasks when they are executed with different partitioning strategies. 
The best partitioning strategy is represented in a dotted bar. The worst partitioning strategy is represented in a diagonally striped bar. 
The best partitioning strategy to execute the \textbf{All-Pair Common Neighborhood} (APCN) algorithm for the \textit{Web-Stanford} graph data is `2D Edge Partition' partitioning strategy while the worst of it is `Hybrid' in Figure \ref{fig1:1}. 
In cases with different algorithms \textbf{PageRank} and \textbf{TriangleCount} for the same graph data, however, the best strategies are `Hybrid' and `Ginger' respectively, and the worst strategies are also different in Figure \ref{fig1:2}, \ref{fig1:4}. 
In addition, the same algorithm APCN and a different graph data \textit{Gemsec-HU} show a different performance order also in Figure \ref{fig1:3}. 
The best partitioning strategy for one task can be the worst strategy for another task as seen in Figure \ref{fig1:1}, \ref{fig1:2}, \ref{fig1:3}, \ref{fig1:5}.


Then, how can we find the most appropriate partitioning strategy that has the best performance for the task? 
We assume that comprehending the graph data and algorithm can help select the best partitioning strategy. 
Several research \cite{10.14778/3055540.3055543, 10.1145/3299869.3300076} compare performances of partitioning strategies and propose a decision tree to select the best partitioning strategy. 
However, they do not declare clear conditions to select decision paths in their decision trees. 
Also, their heuristic decision trees are not appropriate to cover cases with various graph data and algorithms. 
Instead of empirical and heuristic selection of partitioning strategies, we take a machine learning approach that can be generally applied to various graph data and algorithms. 
\cite{10.1145/3301326.3301354, 8729187} considers graph data to select the best partitioning strategy. They chose only one algorithm, PageRank, to compare the performance of partitioning strategies and did not consider algorithm characteristics. 
We instead extract graph data and algorithms' features by carefully analyzing execution behaviors. By that, our method proposes the most suitable strategy to divide data across workers. 
\input{figure/overall5} 
Figure \ref{fig:overall5}-\textcircled{\small{1}}, \textcircled{\small{2}} shows extracting the features of the task. 
The task feature is the concatenation of graph data statistics and algorithm execution pattern features. 
Figure \ref{fig:overall5}-\textcircled{\small{3}} shows predicting the performance for each partitioning strategy using the task feature. 
We used a machine learning technique in this part, and our approach is similar to the concept of software 2.0 supporting systems using data-driven methods. 
There are research papers related to database configuration tuning \cite{10.1145/3299869.3300085, 10.1145/3035918.3064029}, relational table partitioning \cite{10.1145/3318464.3389704} and cardinality estimation \cite{10.14778/3291264.3291267}. 
We select the strategy with the fastest expected execution time in Figure \ref{fig:overall5}-\textcircled{\small{4}}. 
Figure \ref{fig:overall5}-\textcircled{\small{5}} depicts the training process of the Execution Time Regression Model (ETRM). We use the augmented synthetic training dataset as the execution logs and train the model using the loss between these logs and the model's outputs.  

We encounter several challenges in designing and implementing the proposed model. 
First, we have to predict the execution time of a task by extracting its features without actually performing it. 
Next, we need to carefully analyze both algorithms and graph data to extract useful features that can be used for the strategy selection model. 
In addition, we need a large dataset to train the machine learning model. 
Creating a sufficient real execution log for the training dataset consumes much computing power, so we construct a synthetic training dataset by augmenting real execution logs.
Finally, excluding some characteristics of several distributed graph engines, we have to implement an experimental distributed graph engine which the all graph algorithms run on and which covers various partitioning strategies.

We performed several experiments, and a list of experiments is as follows.  \lowercase\expandafter{\romannumeral1}) How well our model can select the best partitioning strategy for test cases, \lowercase\expandafter{\romannumeral2}) how superior the selected strategy's performance is compared to other strategies, and \lowercase\expandafter{\romannumeral3}) how much performance benefit our approach can get. 

The main contributions of our research are the following:
\begin{itemize}
  \item We propose a method to choose the best partitioning strategy using extracted features from graph data and algorithms. 
  \item We construct an experimental distributed graph engine, so only the factors for graph data, algorithm, and partitioning strategy be the experimental elements.
  \item We propose a method to generate synthetic training data to train our model.
\end{itemize}

The rest of this paper is organized as follows. In Section 2
, we summarize notations. 
In Section 3
, we describe our distributed graph computation engine that we implemented for experimental  purposes. 
In Section 4
, we describe the set of features we extracted and how. 
In Section 5
, we evaluate our method. 
In Section 6
, we review related works. 
We conclude and propose future work in section 7
.

%% file: figure/Fig1_motivation.tex
\begin{figure}[!ht]
    \centering
    \begin{subfigure}[b]{0.9\columnwidth}
        \centering
        \includegraphics[width=0.9\textwidth]{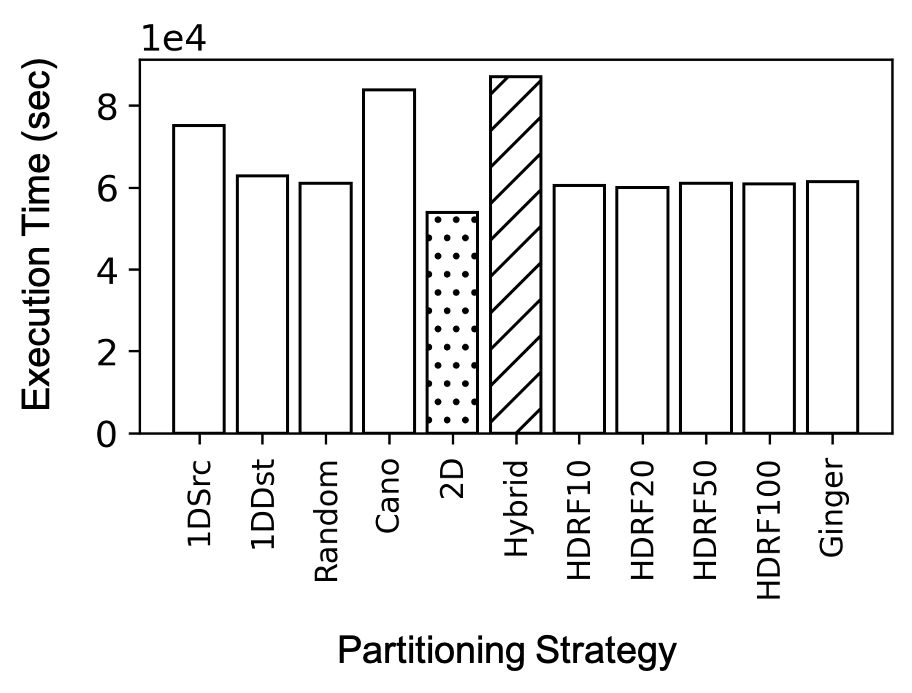}
        \caption{{\textit{Web-Stanford}/\textbf{All-PairCommonNeighbor~ (APCN)}}}    
        \label{fig1:1}
    \end{subfigure}
    \newline
    \begin{subfigure}[b]{0.48\columnwidth}  
        \centering 
        \includegraphics[width=\columnwidth]{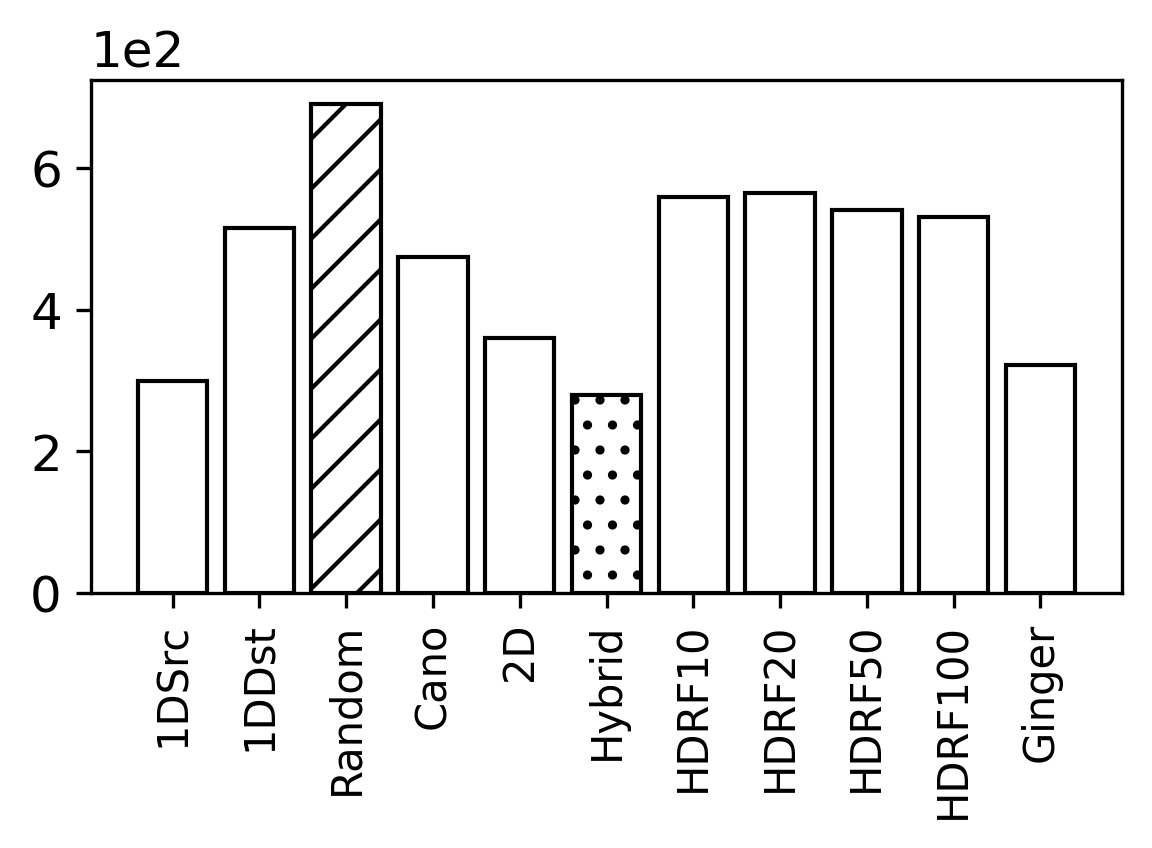}
        \caption[]%
        {{\textit{Web-Stanford}/\textbf{PageRank}}}    
        \label{fig1:2}
    \end{subfigure}
    \hfill
    \begin{subfigure}[b]{0.48\columnwidth}   
        \centering 
        \includegraphics[width=\columnwidth]{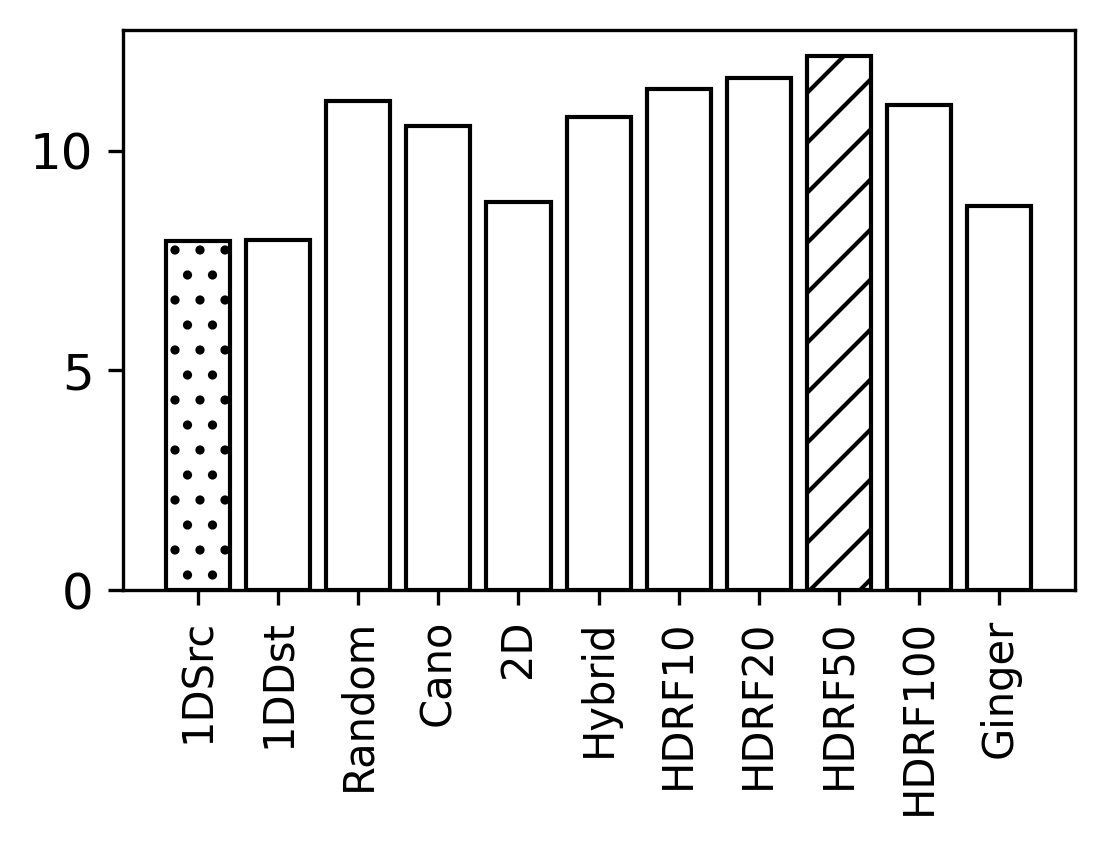}
        \caption[]%
        {{\textit{Gemsec-HU}/\textbf{APCN}}}    
        \label{fig1:3}
    \end{subfigure}
    \newline

    \begin{subfigure}[b]{0.48\columnwidth}
        \centering
        \includegraphics[width=\columnwidth]{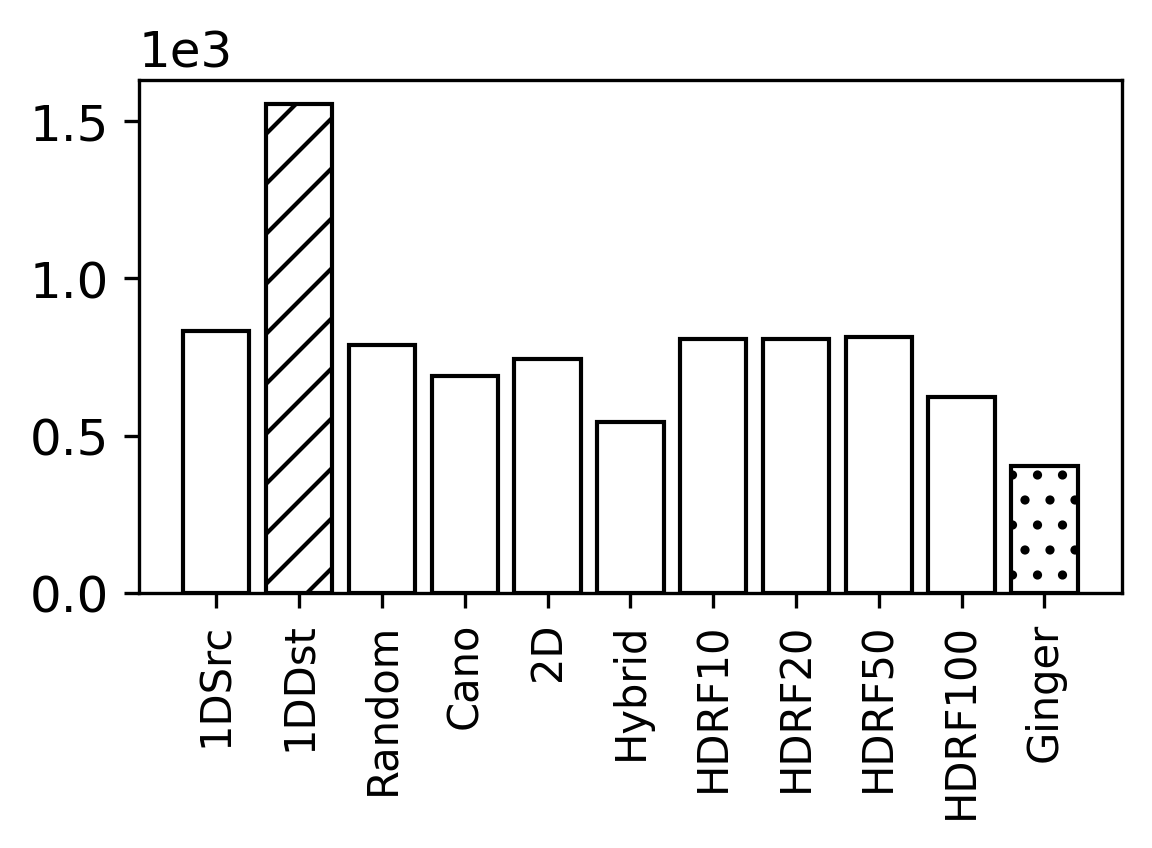}
        \caption[Network2]%
        {{\nolinebreak\textit{Web-Stanford}/\textbf{TriangleCount}}}    
        \label{fig1:4}
    \end{subfigure}
    \hfill
    \begin{subfigure}[b]{0.48\columnwidth}  
        \centering 
        \includegraphics[width=\columnwidth]{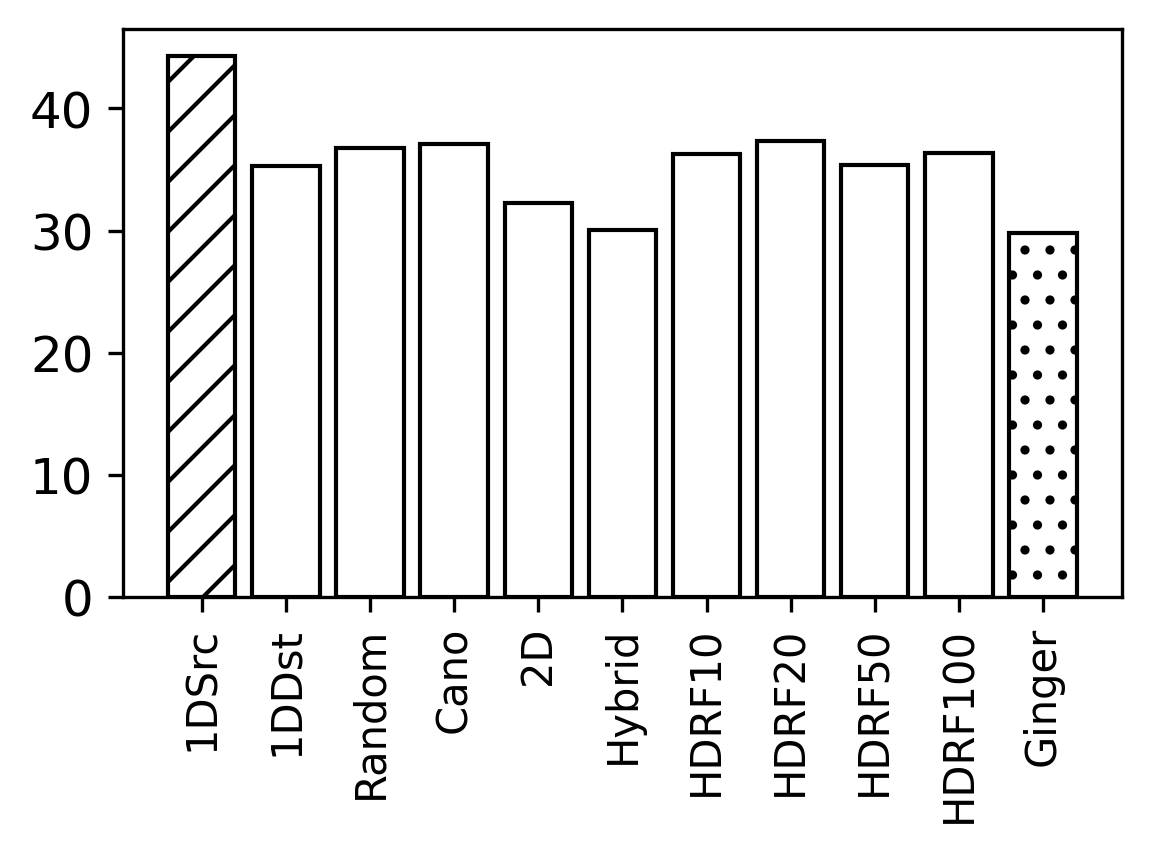}
        \caption[]%
        {{\textit{Gemsec-HR}/\textbf{APCN}}}
        \label{fig1:5}
    \end{subfigure}

    \caption[Performance of Partitioning Strategies]
    { Comparison of the best partitioning strategy, This example is experimented by using our distributed engine mentioned in Section 3.  \textit{Web-Stanford}, \textit{Gemsec-Hu} and \textit{Gemsec-HR} each consist 281,903, 47,538 and 54,573 vertices and 2,312,497, 222,887 and 498,202 edges. This example consists of 64 workers on four identical machines. The specification of one machine is 32 cores, Xeon X7560 2.27GHz, 500GB RAM. Machines communicate using 10 Gbps NICs. }
    \label{fig1:main}
\end{figure}

%% file: figure/overall5.tex
\begin{figure*}[ht]
\begin{center}
  \includegraphics[width=1.6\columnwidth]{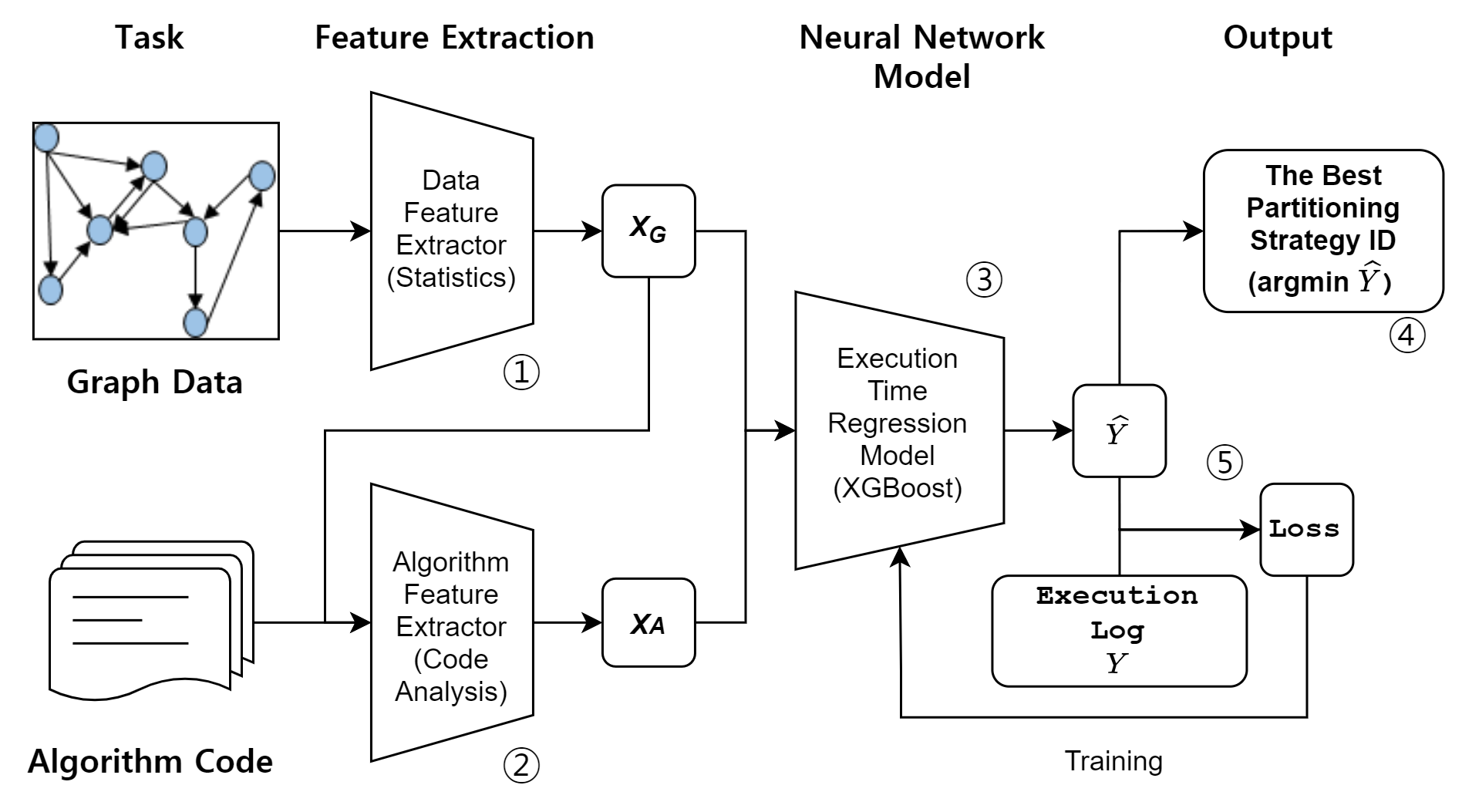}
  \caption{Overall process of finding out the best partitioning strategy}
  \label{fig:overall5}
\end{center}
\vspace{-0.5cm}
\end{figure*}

%% file: 01-05_Symbols.tex
\section{Notations}
\input{table/symbols}
We summarize notations used in this paper in Table \ref{tab:notation}. These notations include the vertex set $V$, edge set $E$, and neighbor vertex $N(u)$ related to the graph data $G$. In addition, Table \ref{tab:notation} includes the worker set $W$, the partitioned vertex set and edge set used in distributed processing, and the notations used in the execution time regression model.

%% file: table/symbols.tex
\begin{table}[t]
\renewcommand\arraystretch{1.5}
\centering
\caption{Notations and descriptions}
\label{tab:notation}
\resizebox{\columnwidth}{!}{%
\begin{tabular}{p{2.5cm}p{5.5cm}}
\toprule
\textbf{Notation} & \textbf{Description} \\
\midrule
$G(V,E)$ & Graph Data $G$ with vertex set $V$ and edge set $E$ \\
$(u,v)$ & An edge whose source vertex $u$ and destination vertex $v$ \\
$N_{in}(v)$, $N_{out}(v)$ & In/Out-neighbor vertices of the vertex $v$ \\
$Degree(v)$ & The number of edges that are incident to the vertex $v$ \\
$W$ & Workers $W = \{w_{1}, w_{2}, ..., w_{i}\}$ \\
$V_{w_{x}}$, $E_{w_{x}}$ & Vertices/Edges of the worker $w_{x}$ \\
$X_{G}$, $X_{A}$ & Graph data / Algorithm features \\
$P$ & Partitioning strategy list $P = \{p_{1}, p_{2}, ..., p_{j}\}$ \\
$\widehat{y_{p_{j}}}$ & Task's expected execution time with a partitioning strategy $p_{j}$ \\
$\widehat Y$ & $\widehat{Y} = \{\widehat{y_{p_{1}}}, ..., \widehat{y_{p_{|P|}}}\}$ \\
$y_{p_{j}}$ & Task's real execution time with a partitioning strategy $p_{j}$ \\
$Y$ & $Y = \{y_{p_{1}}, ..., y_{p_{|P|}}\}$ \\
\bottomrule
\end{tabular}%
}
\end{table}

%% file: 03_distributed_graph_engine.tex
\section{Distributed Graph Engine}
\input{figure/distributed_graph}
This section describes our graph computation engine, which serves as a test bed for comparing the partitioning strategy execution process. This paper focused on the partitioning strategies related to the data and the algorithm. Therefore, our graph engine contains essential functions and operators that serve our purpose. 

\subsection{Graph Representation}
As we focused only on the task's performance, we simplified the implementation. 
Our graph engine used an edge list to represent the graph data. 
The edge list consists of vertex tuples, ${(u, v)}$. 
An inverted edge list is also maintained. 
Finding a vertex takes $O(log(|V|))$ time. 
It takes $O(degree(v))$ to search for an edge connected to an arbitrary vertex $v$ by managing a key-value hash map with vertex id as a key and the starting point of the edge list connected to this vertex as value. 
The edge list is sorted by source vertex ID. 
Thus, insertion and deletion are also ignored. 
In addition, vertex and edge properties are stored in each key-value map. 

\subsection{Distributed Computation Model}

\subsubsection{GAS Model for Distributed Computing}
Among several distributed graph computation models, we selected the GAS model\cite{10.5555/2387880.2387883}. 
Hadoop MapReduce\cite{10.1109/MSST.2010.5496972} is general and can be used in various applications, but is not suitable because it uses HDFS\cite{borthakur2007hadoop}, which can cause excessive I/O and it may run unnecessary shuffle operations. 
TUX$^2$\cite{10.5555/3154630.3154684} proposed the MEGA model, which is optimized for graph machine learning algorithms. 
We didn't adopt the MEGA model because we target more general graph processing tasks instead of specific graph ML tasks.

The GAS model is a vertex-centric model\cite{mccune2015thinking} and `GAS' stands for Gather, Apply and Scatter.
While partitioning edges separately, vertices that exist commonly in partitioned edges are replicated. 
The GAS model sets one vertex as the master vertex and the others as mirror vertices with the same \textbf{vertex ID} vertices.
Workers have a queue for representing vertices that will be processed locally.
Each worker pops a vertex from the queue and propagates it to the corresponding workers having mirror vertices.
For each of these vertices in the Gather phase, the engine collects all mirrors' local results and aggregates them.
In the Apply phase, the master vertex's aggregated result is transmitted to mirror vertices.
In the Scatter phase, the vertex's aggregated result is used to update its adjacent edges. 
The neighbor vertices are enqueued if these neighbor vertices are needed to be computed. 
This activation occurs based on the local neighbor, and this result is shared between workers. 
Vertex $3$'s GAS step is illustrated in Figure \ref{fig:distributed_graph}. In this example, $v_{3}$'s partial result is computed in each worker and aggregated to worker 0. Then, this aggregated value is updated to mirror vertices. Lastly, in this example, out-neighbor $v_{5}$ is activated and en-queued.


\subsubsection{Scalability}
\input{figure/scalability}
We tested the scalability of our engine to show that the implementation is scalable enough to conduct our experiments. 
This result can be seen in Figure \ref{fig:scalability}. 
This experiment consisted of 4, 8, 16, 32, and 64 workers on four identical machines. The specification of one machine is 32 cores, Xeon X7560 2.27GHz, 500GB RAM. Machines communicate using 10 Gbps NICs. 
\textbf{PageRank} and \textbf{TriangleCount} algorithms were performed for \textit{Web-Stanford} data. 
We could see that execution time decreased for two algorithms up to 64 workers. 


\subsection{Partitioning Method}
\input{table/partitioning_strategies}

The partitioning methods used in our test bed were selected based on the following criteria: \lowercase\expandafter{\romannumeral1}) commonly used in many systems, and \lowercase\expandafter{\romannumeral2}) proper to processing model. 
We selected GAS as the distributed graph processing model, and accordingly, we employed partitioning methods supported by representative GAS systems such as GraphX, PowerGraph, and PowerLyra. 
The following describes the partitioning methods supported by our engine. 
Table \ref{table:partition_strategies} shows a brief summary of each partitioning strategy. 
\subsubsection{GraphX}
\renewcommand{\labelenumi}{\roman{enumi}}
\begin{enumerate}
    \item 1D Edge Partition: Hashing is performed based on the ID of the edge's source vertex \textbf{u}. All edges with the same source are mapped to the same worker.


    \item Random: Both \textbf{u} and \textbf{v}'s IDs are input to the hash function, but the reversed order IDs do not necessarily output the same mapping result. Cantor pairing function\cite{lisi2007some} can be used to map a 2D input to a 1D output.
    \item Canonical Random: This method is similar to Graph X's Random method. The ordered \textbf{u} and \textbf{v} IDs are applied as a hash function input. Both (\textbf{u}, \textbf{v}) edge and (\textbf{v}, \textbf{u}) edge are mapped to the same worker.

    \item 2D Edge Partition: This method applies hashing to the source vertex \textbf{u} and destination vertex \textbf{v} of the edge, respectively. Two-dimensional mapping is performed, and worker IDs are assigned to each two-dimensional tile. Moreover, in this partitioning strategy, when \textbf{$|W|$} is a square number, each vertex must have a maximum of $2\sqrt{\textbf{$|W|$}}$ replicated vertices \cite{10.1145/2484425.2484429}. 

%

\end{enumerate}
\subsubsection{PowerGraph}
\renewcommand{\labelenumi}{\roman{enumi}}
\begin{enumerate}
    \item Random: PowerGraph's Random method is similar to Graph X's Canonical Random method. The worker ID that will contain the edge is selected by hashing the input of the two vertices of the edge, regardless of the edge direction.
    
    \item Greedy Vertex-Cuts (Oblivious): This method is not based on a hash. Edges are distributed one by one, successively using a greedy method. The Oblivious method checks the edge distribution condition and assigns an edge to a worker to have the number of replicated vertices as little as possible and balance the number of allocated edges. A detailed explanation can be found in \cite{10.5555/2387880.2387883}. However, we observed this method sometimes fails to utilize all workers. Hence, we exclude this from our inventory of strategies. 
    
    \item HDRF: This method is the abbreviation for High-Degree Replicated First\cite{10.1145/2806416.2806424} which Petoni et al proposed. HDRF measures the partition score of each worker for new edge assignments. In the replication term ${{C}_{REP}}$, if worker \textbf{$w_{x}$} already contains a vertex belonging to a new edge, a score will be added. If a score will be added, the lower the partial degree rate of the vertex, the higher the score. In the balance term ${{C}_{BAL}}$, the smaller the number of edges belonging to the worker, the higher the score. In this way, HDRF can maintain a low replication factor and effectively distribute the load. In this paper, we tested by changing the values for $\lambda$ to 10, 20, 50, 100.
    
    {\small
    \begin{align}
        {Score}(u, v, w_{x})={{C}_{REP}}(u, v, w_{x})+\lambda*{{C}_{BAL}}(w_{x})
    \end{align}
    }%
    
\end{enumerate}

\subsubsection{PowerLyra}
\renewcommand{\labelenumi}{\roman{enumi}}
\begin{enumerate}
    \item Hybrid: Hybrid is a method that considers neighbor-related functions applied to graphs. This method allocates neighboring vertices to the same worker. For load-balancing, not all neighbor vertices are assigned to one worker. PowerLyra sets a threshold for the number of neighbors, and the allocation method is different depending on the neighbor's degree. For example, when calculating the PageRank\cite{page1999pagerank}, the PageRank of vertex $\alpha$ is the sum of the vertex values connected by the in-edge of $\alpha$. If $\alpha$ has an in-neighbor degree below the threshold, all in-neighbor vertices of $\alpha$ are assigned to the same worker. Otherwise, the edges connected to these vertices are partitioned to workers, not to only one worker, using hashing to the edge's source vertex. 
    
    \item Ginger: The Ginger method is similar to the hybrid method. The hybrid method divides the graph using hashing and vertex degree, but Ginger has one more process of calculating each vertex's score for each worker. To the worker with the highest score, the target vertex and its in-neighbor vertices are partitioned. Ginger's score function is as follows.
    
    {\footnotesize
    \begin{align}
        {Ginger}(v, w_{x}) = |{{N}_{in}}(v) \cap {{V}_{w_{x}}}|
        - \frac{1}{2}(|{{V}_{w_{x}}}|+\frac{|V|}{|E|}|{{E}_{w_{x}}}|)
    \end{align}
    }%
    
    The larger the intersection $|{{N}_{in}}(v) \cap {{V}_{w_{x}}}|$, the more the number of in-neighbor vertices of $v$ that worker \textbf{$w_{x}$} holds. This term suppresses the increase in the replication factor. For the second term $\frac{1}{2}(|{{V}_{w_{x}}}|+\frac{|V|}{|E|}|{{E}_{w_{x}}}|)$, the smaller the vertices and edges included in the worker, the higher the score. This helps balance the load between workers. 
    
\end{enumerate}

\subsubsection{Others}
We include our custom partitioning strategy. We call it the 1D Edge Partition-Destination method, which is similar to Graph X's 1D Edge Partition method. This strategy divides edges based on the ID of the edge's destination vertex \textbf{$v$}

The edge-cut method can assure load-balance in scenarios dealing with large-scale power-law real-world graphs. For example, Fennel\cite{10.1145/2556195.2556213} can reduce edge-replication than hash-based strategy in some graph data. However, it has a high communication cost because it forms a high replication ratio, edge-cut ratio, for the edge\cite{10.14778/3055540.3055543}. For graph data with small and non-distorted distributions, hybrid and Ginger strategy work as an edge-cut strategy. Therefore, our contents do not cover the edge-cut strategies.


%% file: figure/distributed_graph.tex
\begin{figure}[ht]
\begin{center}
  \includegraphics[width=0.9\columnwidth]{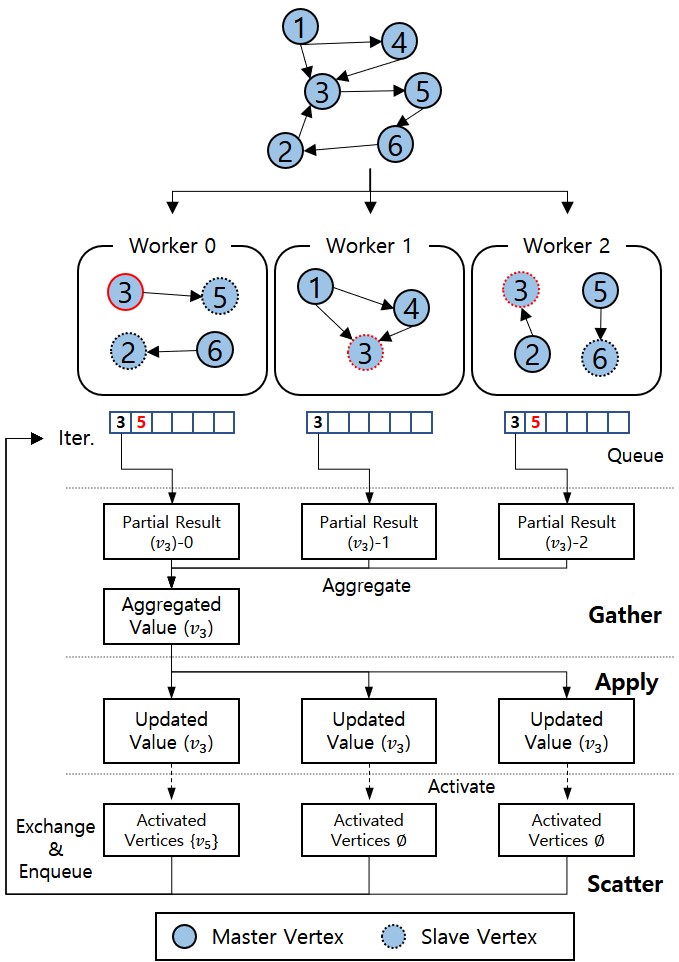}
  \caption{Example of partitioned graph $\And$ graph computation
  }
  \label{fig:distributed_graph}
\end{center}
\vspace{-0.5cm}
\end{figure}

%% file: figure/scalability.tex
\begin{figure}[t]
    \centering
    \begin{subfigure}[b]{0.475\columnwidth}
        \centering
        \includegraphics[width=\textwidth]{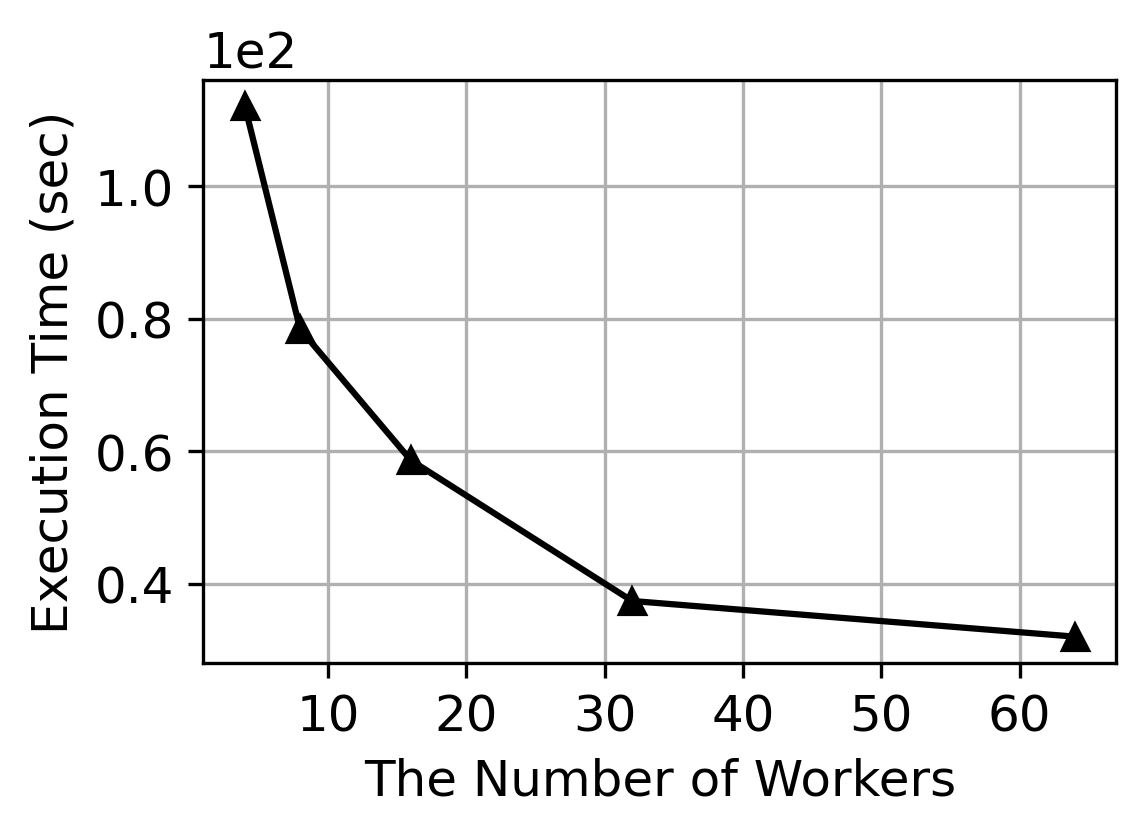}
        \caption[Network2]%
        {{\small \textbf{PageRank}, 10 iterations}}    
        \label{scal-PR}
    \end{subfigure}
    \hfill
    \begin{subfigure}[b]{0.475\columnwidth}  
        \centering 
        \includegraphics[width=\textwidth]{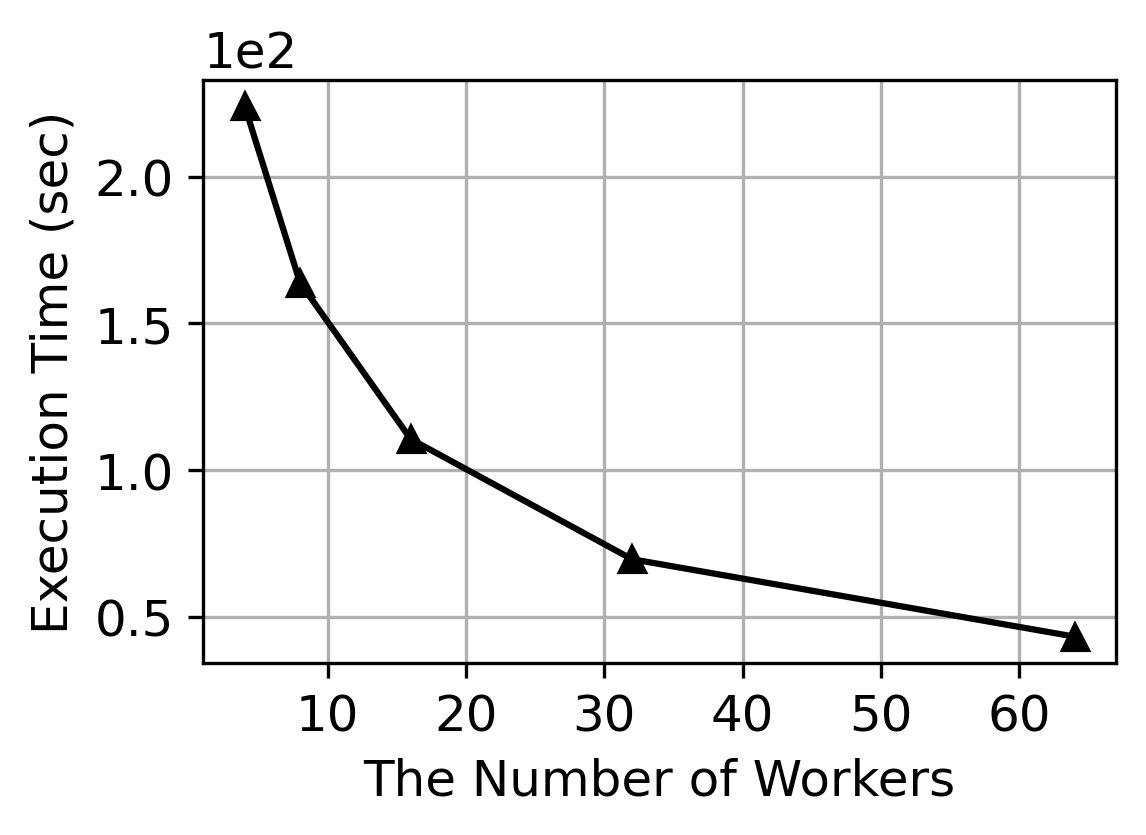}
        \caption[]%
        {{\small \textbf{TriangleCount}}}    
        \label{scal-TC}
    \end{subfigure}
    
    \caption[]
    {\small Engine scalability test on the \textit{Web-Stanford} graphs with \emph{2D} partitioning strategy and increasing workers.} 
    \label{fig:scalability}
\end{figure}


%% file: table/partitioning_strategies.tex
\begin{table*}[ht]
  \centering
  \caption{Partitioning Strategies}
  \label{table:partition_strategies}
  \begin{tabular}{m{0.5cm}m{4cm}m{2.4cm}m{3.3cm}m{3.0cm}m{1.7cm}}
    \toprule
    PSID & Strategy Name & Supporting Engine & Method & Target Objects & Remarks\\
    
    \midrule
    0 & 1D Edge Partition (\emph{1DSrc})  & Graph X    & 1D-Hash                  & -     & -                           \\ 
    1 & 1D Edge Partition-DST (\emph{1DDst})       & -  & 1D-Hash    & -   & - \\ 
    2 & Random (\emph{Random})             & Graph X    & 2D-Hash                  & -      & -                        \\ 
    3 & Canonical Random (\emph{Cano})   & Graph X    & 2D-Hash                  & -       & -                         \\ 
    4 & 2D Edge Partition (\emph{2D})  & Graph X    & Two 1D-Hash              & -      & -                    \\ 
    5 & Hybrid (\emph{Hybrid})             & PowerLyra  & Hash \& Degree threshold & Replication Factor   & -          \\ 
    6 & Greedy Vertex-Cuts (\emph{Oblivious}) & PowerGraph & Greedy                   & Replication Factor     & -        \\ 
    7-10 & HDRF (\emph{HDRF})      & PowerGraph & Greedy                   & Replication Factor \& Load-balance  & Lambda values: 10, 20, 50, 100 \\  
    11 & Ginger (\emph{Ginger})             & PowerLyra  & Greedy                   & Replication Factor \& Load-balance  & - \\
  \bottomrule
\end{tabular}
\end{table*}

%% file: 04_my_model.tex
\section{Static Analysis and Evaluation} \label{methodology}
To prove our hypothesis that we can choose a better partitioning strategy by analyzing the data and the algorithm, we extracted certain features. The structural properties of graph data are summarized by statistic values, and symbolic code analysis on the pseudo-code of the algorithm is conducted to get the algorithm features. Finally, a machine learning model is constructed to predict how long does it take for given tasks.
The overall process can be seen in Figure \ref{fig:overall5}. Section \ref{feature_extractor} explains how the features are extracted from the graph data and algorithm. Section \ref{regression_model} explains the Execution Time Regression Model that predicts the performance of partitioning strategy using machine learning. 


\subsection{Feature Extractor} \label{feature_extractor}

\subsubsection{Data Feature}
\input{table/data_feature}


Various inherent features in the graph data were selected for the following reasons and summarized in Table \ref{tab:data_features}. 

The number of vertices and edges is helpful in analyzing the iteration over the entire graph and predicting the iteration's execution time. 

Graph data always represent the relationship between vertices as edges, and graph data analysis commonly accompanies access to edges. 
Therefore, it is necessary to understand the graph topology because the degree of vertices and their distribution vary according to the topology. 
We extracted mean, standard deviation, skewness, and kurtosis from each vertices' in-degree and out-degree. Since skewness and kurtosis can have negative values, they are divided into a sign and absolute value and used as input features. 

Furthermore, it is essential to consider whether the graph is directed because some operators behave differently (e.g. get in-neighbors of some vertex, inverted edge list). 


\subsubsection{Algorithm Feature}
\input{table/application_feature}

The frequency of graph operations are evaluated to capture the pattern and scale of data access for executing the graph algorithm. 
We wrote a code analyzer with \textit{JavaCC}, a compiler-compiler tool like \textit{YACC}. It analyzes the pseudo-code which consists of graph operators supported by our engine. The operators are listed in Table \ref{tab:algo_features}.



\input{listing/01_pagerank}
\input{listing/02_counts}

The pseudo-code has symbols representing graph elements such as \texttt{ALL\_VERTEX\_LIST, GET\_IN\_VERTEX\_TO} as seen in Listing \ref{list:pagerank}.
Parsing the code, the number of each graph and arithmetic operation is counted. 
As a result, the key-value pairs as operation-count will be generated as seen in line 1 of Listing \ref{list:count}. 
Even if it cannot be evaluated as a real value, the count is represented by a symbolic expression. 
In order to fill in those symbols with real values, data features are used to evaluate the symbols. 
For example, a graph operation \texttt{GET\_IN\_VERTEX\_TO(v)} in line 10 of Listing \ref{list:pagerank} is found at the condition of for loop in line 10. The number of this operation can be evaluated by the multiplication of outer loop variables, \texttt{|ALL\_VERTEX\_LIST|}*\texttt{iterator\_num}. 
The number of iteration can be found in line 1 of Listing \ref{list:pagerank}, so it is immediately evaluated as 20.0. 
The size of all vertex set is trivially same as $|V|$, the cardinality of the vertex set of graph. Thus, the number of all vertex can be taken from data feature $DF$. The graph data in this example was Ego-Facebook\cite{snapnets} and $|V|$ is $4039$ so, the final counting value of \texttt{GET\_IN\_VERTEX\_TO} becomes $4039*20.0 = 80780.0$.


The accesses to variables and arithmetic operators are also counted to precisely evaluate the loop's body i.e. line 5, 12 in the Listing \ref{list:count}.


\subsection{Execution Time Regression Model} \label{regression_model}
We implemented a prediction model to select the best partitioning strategy in a given task. We tried some machine learning models such as linear regression, XGBoost\cite{10.1145/2939672.2939785},  LightGBM\cite{10.5555/3294996.3295074}, multi-layer perceptron and mixture of experts \cite{doi:10.1162/neco.1994.6.2.181}. 
The best model was the XGBoost regression model.
The training process and model structure of the Execution Time Regression Model are as follows.

\subsubsection{Data Preparation}
We executed tasks encompassing the graph data, algorithms, and partitioning strategies on our engine and recorded execution logs.
The graph data and algorithm in the execution log are mapped to the data feature and the algorithm feature respectively by the feature extractor.
We prepared total execution logs using 12 graph data, 8 algorithms, and 11 partitioning strategies except \textit{Oblivious} strategy. 

To train our machine learning model, we had to prepare a huge amount of training set. 
Among those execution logs, 528 logs made by 8 graph data, 6 algorithms, and 11 partitioning strategies were used to create the augmented training dataset. 
We generated synthetic data via aggregation of multiple real data records. 
Aggregation is the summation of the algorithm feature and execution time by grouping the logs performed in the same graph data with the same partitioning strategy. 
The task of a synthetic tuple is interpreted as one large algorithm with several algorithms performed sequentially. 
Therefore, the algorithm features that predict the number of calls for the low-level function and the execution time can be aggregated by summation. 
For example, if a synthetic tuple $s$ is created via aggregation of real tuples $r_{1}, r_{2}, ..., r_{n}$, then tuple $s$'s algorithm feature is $AF(s) = \sum_{i}^{n} AF(r_{i}) $. 
It's data feature is $DF(s) = DF(r_{1}) = ... = DF(r_{n})$ and execution time is $ET(s) = \sum_{i}^{n} ET(r_{i})$. 

We used combinations with replacement to make the synthetic algorithms. 
The formula is as follows. 
\begin{align}
C^{R}(n, r)= \frac{(n+r-1)!}{r!(n-1)!}
\end{align}
We created synthetic algorithms by using 6 original algorithms and changing r from 2 to 9. 
The number of synthetic algorithms is $\sum_{r=2}^{9} C^{R}(6, r) = 4998$. 
Our synthetic training dataset has about 0.43 million tuples by multiplying 4998 synthetic algorithms, 8 graph data and 11 partitioning strategies. 
Since each sample of the synthetic dataset has a different combination of algorithms, the entire synthetic dataset can be interpreted as a record of performing various and unique algorithms. 
The augmented training dataset does not include the original 528 real records. 
We used 528 records and records from other 4 graph data and 2 algorithms in the test phase. 



\subsubsection{Model Architecture}
Our model captures the features of data and an algorithm to predict their execution time of a task for a given partitioning strategy. 
XGBoost regression model includes regularization term and uses Classification And Regression Tree (CART) ensemble model. 
This ensemble model decides whether to split the branch to maximize the Gain and minimize the Objective function. 
The formula is as follows. 
\begin{align}
&\hat{y_{i}}^{(t)} = \text{prediction of the i-th instance at the t-th iteration} \\
&loss(y_{i}, \hat{y_{i}}^{(t-1)}) = (y_{i} - \hat{y_{i}}^{(t-1)})^{2} \\
&g_{i} = \partial_{\hat{y_{i}}^{(t-1)}} loss(y_{i}, \hat{y_{i}}^{(t-1)}) \\
&h_{i} = \partial^{2}_{\hat{y_{i}}^{(t-1)}} loss(y_{i}, \hat{y_{i}}^{(t-1)}) \\
&I_{L} = \text{the instance sets of left node after the split} \\
&I_{R} = \text{the instance sets of right node after the split} \\
&I = I_{L} \cup I_{R} \\
&\lambda = \text{L2 regularization term on weights} \\
&\gamma = \substack{\textstyle \text{minimum loss reduction required to make a} \\ \textstyle \text{further partition on a leaf node of the tree}} \\
&Gain = \frac{(\sum_{i\in I_{L}}g_{i})^{2}}{\sum_{i\in  I_{L}}h_{i}+\lambda } + \frac{(\sum_{i\in I_{R}}g_{i})^{2}}{\sum_{i\in I_{R}}h_{i}+\lambda } - \frac{(\sum_{i\in I}g_{i})^{2}}{\sum_{i\in I}h_{i}+\lambda } - \gamma \\
&\Omega = \text{regularization term} \\
&f_{k} = \text{k-th decision tree in function space F} \\
&objective = \sum_{i=1}^{|Samples|} loss(y_{i}, \hat{y_{i}}^{(t-1)}) + \sum_{k=1}^{K} \Omega (f_{k})
\end{align}

We used the XGBRegressor model and the detailed parameters of the Regressor are as follows.
\begin{itemize}
     \item colsample\_bytree = 0.4603
     \item gamma = 0.0468
     \item learning\_rate = 0.05
     \item max\_depth = 15
     \item min\_child\_weight = 1.7817
     \item n\_estimators = 1000
     \item reg\_alpha = 0.4640
     \item reg\_lambda = 0.8571
     \item subsample = 0.5213
     \item objective = squared error
\end{itemize}
The input of the model is expressed as $X$. 
\input{figure/encoding} $X$ has both graph data features $X_{G}$ and algorithm features $X_{A}$ pre-processed with scaling and one-hot encoding. 
Figure \ref{fig:encoding_info} describes the model's input data. 

%% file: table/data_feature.tex
\begin{table}[b]
\renewcommand\arraystretch{1.1}
\centering
\caption{Data features}
\label{tab:data_features}
\resizebox{\columnwidth}{!}{%
\begin{tabular}{ccccc}
    \toprule
    Category & Feature & Type & \begin{tabular}[c]{@{}c@{}}Gain\\ Importance\end{tabular} & \begin{tabular}[c]{@{}c@{}}Split\\ Importance\end{tabular} \\
    \midrule
    \multirow{2}{*}{Cardinality} & The number of Vertex & Int & 0.1855 & 1211 \\
 & The number of Edge & Int & 0.2472 & 859 \\
    \hdashline
    \multirow{2}{*}{Topology} & In-degree & Float & 0.0578 & 528 \\
 & Out-degree & Float & 0.4326 & 1368 \\
    \hdashline
    Direction & Graph direction & Categorical & 0.0046 & 119 \\
    \bottomrule
\end{tabular}%
}
\end{table}


%% file: table/application_feature.tex

\begin{table}[t]
\centering
\caption{Algorithm features}
\label{tab:algo_features}
\resizebox{\columnwidth}{!}{%
\begin{tabular}{ccccc}
\toprule
Category & Feature & Description & \begin{tabular}[c]{@{}c@{}}Gain\\ Importance\end{tabular} & \begin{tabular}[c]{@{}c@{}}Split\\ Importance\end{tabular} \\
\midrule
\multirow{5}{*}{\begin{tabular}[c]{@{}c@{}}Graph\\ Object\end{tabular}} & NUM\_VERTEX & \begin{tabular}[c]{@{}c@{}}Get the number \\ of all vertices\end{tabular} & 0.0048 & 654 \\
 & NUM\_EDGE & \begin{tabular}[c]{@{}c@{}}Get the number \\ of all edges\end{tabular} & - & - \\
 & NUM\_IN\_DEGREE & \begin{tabular}[c]{@{}c@{}}Get in-degree \\ of a vertex\end{tabular} & - & - \\
 & NUM\_OUT\_DEGREE & \begin{tabular}[c]{@{}c@{}}Get out-degree \\ of a vertex\end{tabular} & 0.0026 & 1101 \\
 & NUM\_BOTH\_DEGREE & \begin{tabular}[c]{@{}c@{}}Get degree \\ of a vertex\end{tabular} & - & - \\
 \hdashline
\multirow{5}{*}{\begin{tabular}[c]{@{}c@{}}Graph\\ Iteration\end{tabular}} & ALL\_VERTEX\_LIST & Set of all vertices & 0.0100 & 1972 \\
 & ALL\_EDGE\_LIST & Set of all edges & 0.0007 & 1755 \\
 & GET\_IN\_VERTEX\_TO & \begin{tabular}[c]{@{}c@{}}In-neighbors \\ of a vertex\end{tabular} & 0.0056 & 2790 \\
 & GET\_OUT\_VERTEX\_FROM & \begin{tabular}[c]{@{}c@{}}Out-neighbors \\ of a vertex\end{tabular} & 0.0020 & 3386 \\
 & GET\_BOTH\_VERTEX\_OF & \begin{tabular}[c]{@{}c@{}}Neighbors \\ of a vertex\end{tabular} & 0.0032 & 2206 \\
 \hdashline
\multirow{4}{*}{\begin{tabular}[c]{@{}c@{}}Graph\\ Operation\end{tabular}} & VERTEX\_VALUE\_READ & \begin{tabular}[c]{@{}c@{}}Read operation \\ on vertex type\end{tabular} & 0.0123 & 2962 \\
 & VERTEX\_VALUE\_WRITE & \begin{tabular}[c]{@{}c@{}}Write operation \\ on vertex type\end{tabular} & 0.0018 & 3648 \\
 & EDGE\_VALUE\_READ & \begin{tabular}[c]{@{}c@{}}Read operation \\ on edge type\end{tabular} & - & - \\
 & EDGE\_VALUE\_WRITE & \begin{tabular}[c]{@{}c@{}}Write operation \\ on edge type\end{tabular} & - & - \\
 \hdashline
\multirow{7}{*}{Basic} & ADD & Add operation & 0.0020 & 2459 \\
 & SUBTRACT & Subtract operation & 0.0061 & 5200 \\
 & MULTIPLY & Multiply operation & 0.0126 & 404 \\
 & DIVIDE & Divide operation & 0.0022 & 1207 \\
 & OTHERS\_VALUE\_READ & \begin{tabular}[c]{@{}c@{}}Read operation \\ on other type\end{tabular} & - & - \\
 & OTHERS\_VALUE\_WRITE & \begin{tabular}[c]{@{}c@{}}Write operation \\ on other type\end{tabular} & 0.0040 & 3165 \\
 & APPLY & \begin{tabular}[c]{@{}c@{}}Apply operation \\ on vertices\end{tabular} & 0.0018 & 2825 \\
 \bottomrule
\end{tabular}%
}
\end{table}

%% file: listing/01_pagerank.tex
\definecolor{mygreen}{rgb}{0,0.6,0}
\definecolor{mygray}{rgb}{0.5,0.5,0.5}
\definecolor{mymauve}{rgb}{0.58,0,0.82}
\lstset{ 
  language=C,
  backgroundcolor=\color{white},   
  basicstyle=\ttfamily\footnotesize,
  breakatwhitespace=false,         
  breaklines=true,                 
  captionpos=b,                    
  commentstyle=\color{mygreen},    
  deletekeywords={...},            
  escapeinside={\%*}{*)},          
  extendedchars=true,              
  firstnumber=1,                
  keepspaces=true,                 
  keywordstyle=\bfseries\color{blue},
  morekeywords={*,...},            
  numbers=left,                    
  numbersep=5pt,                   
  numberstyle=\ttfamily\color{mygray},
  showspaces=false,
  showstringspaces=false,          
  showtabs=false,                  
  stepnumber=1,                    
  stringstyle=\color{mymauve},
  tabsize=2,
  title=\lstname,
  caption=Code analysis example: PageRank,
  xleftmargin=5.0ex
}

\begin{lstlisting}[label={list:pagerank}, float]
int iterator_num = 20;
float dampling_factor = 0.85;
float temp_value;
for(list v in ALL_VERTEX_LIST){
    v.value = 1.0 / NUM_VERTEX;
}
for(iterator_num){
	for(list v in ALL_VERTEX_LIST){
		temp_value = 0;
		for(list v_in in GET_IN_VERTEX_TO(v)){
			temp_value = temp_value + v_in.value/v_in.NUM_OUT_DEGREE;
		}
		v.value = (1-dampling_factor)/NUM_VERTEX + dampling_factor * temp_value;
		Global.apply(v, "float");
	}	
}
\end{lstlisting}

%% file: listing/02_counts.tex
\definecolor{mygreen}{rgb}{0,0.6,0}
\definecolor{mygray}{rgb}{0.5,0.5,0.5}
\definecolor{mymauve}{rgb}{0.58,0,0.82}
\lstset{ 
  language=Python,
  backgroundcolor=\color{white},   
  basicstyle=\ttfamily\footnotesize,
  breakatwhitespace=false,         
  breaklines=true,                 
  captionpos=b,                    
  commentstyle=\color{mygreen},    
  deletekeywords={...},            
  escapeinside={\%*}{*)},          
  extendedchars=true,              
  firstnumber=1,                
  keepspaces=true,                 
  keywordstyle=\bfseries\color{blue},
  morekeywords={*,...},            
  numbers=left,                    
  numbersep=5pt,                   
  numberstyle=\ttfamily\color{mygray},
  showspaces=false,
  showstringspaces=false,          
  showtabs=false,                  
  stepnumber=1,                    
  stringstyle=\color{mymauve},
  tabsize=2,
  title=\lstname,
  caption=Example of counting operations,
  xleftmargin=5.0ex
}

\begin{lstlisting}[label={list:count}, float]
IR = {
	'get_in_vertex_to': AllOfPartSetV*20.0,
	'all_vertex_list': 20.0 + 1,
	'out_edge_num': InVertexSetToPartOfAllV*AllOfPartSetV*20.0,
	'vertex_value_read': InVertexSetToPartOfAllV*AllOfPartSetV*20.0,
    ...
}
Eval = {
    'get_in_vertex_to': 80780.0,
    'all_vertex_list': 21.0,
    'out_edge_num': 3529358.98,
    'vertex_value_read', 3529358.98,
    ...
}

\end{lstlisting}
%

%
    

%% file: figure/encoding.tex
\begin{figure}[t]
\begin{center}
  \includegraphics[width=0.8\columnwidth]{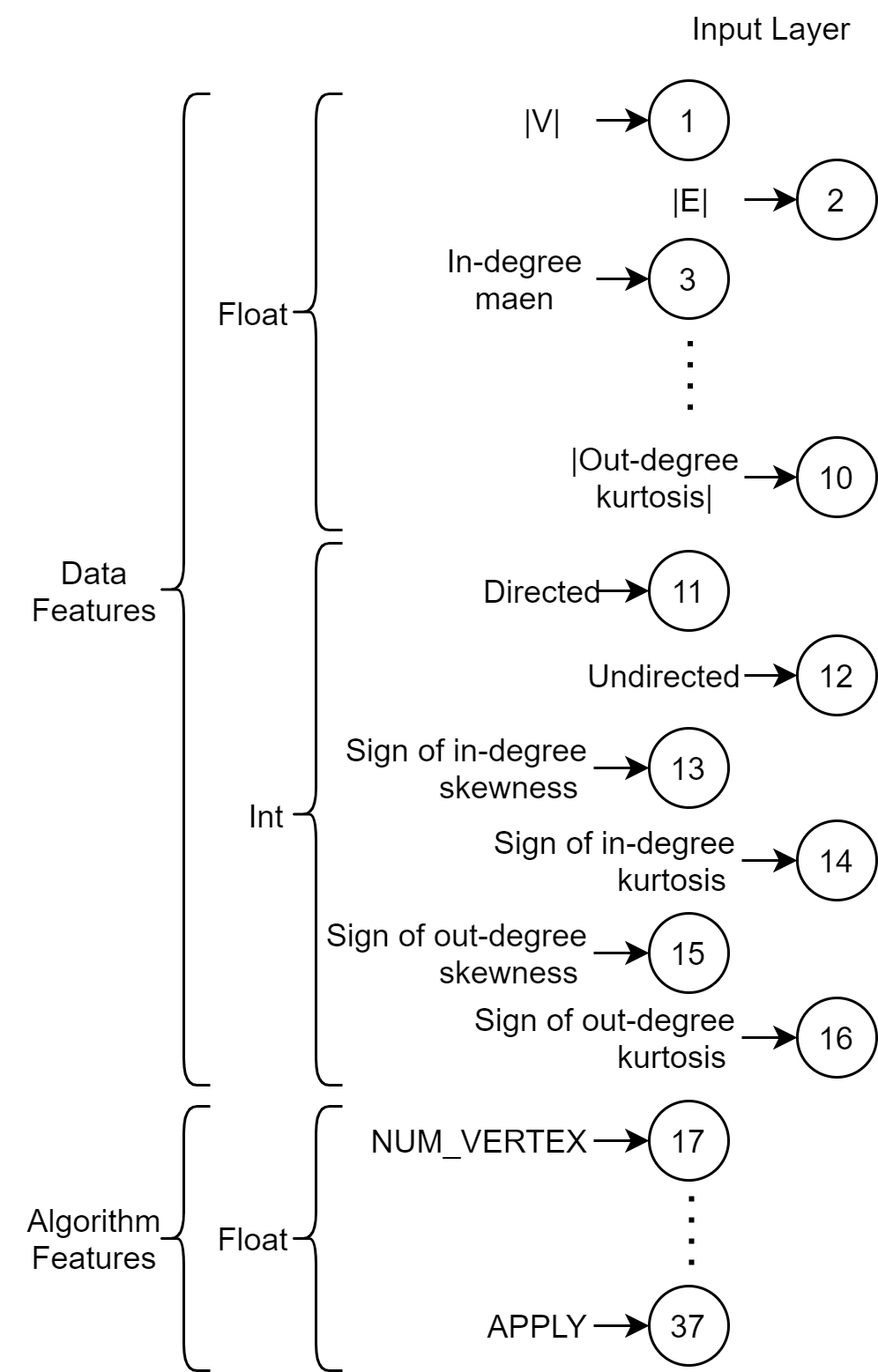}
  \caption{Encoding of input data}
  \label{fig:encoding_info}
\end{center}
\vspace{-0.5cm}
\end{figure}

%% file: 05_experiment.tex
\section{Experiments}
\subsection{Experimental Setup}
We ran our experiments on 64 workers in a local cluster. The cluster consists of 4 same machines, and each node has 16 workers as processes. 
Each worker communicates using Message Passing Interface\cite{gabriel2004open}. 
Hardware Specifications of a machine are Xeon 2.27GHz 32 cores CPU and 503GB RAM. 
The data exchange bandwidth between all workers is 10Gbps. 
Heterogeneous hardware systems are not considered. 
\subsection{Graph Data}
\input{table/dataset2}

For experiments, we used graph data collected in various fields with diverse sizes and topology.
A summary of the graph data is provided in Table \ref{tab:dataset_table2}. All graph data were provided by Stanford Network Analysis Project (SNAP) \cite{snapnets}. The Gemsec Deezer data and Web-Stanford data were never used in creating the augmented training dataset. 
Instead, these graph data are used in model evaluation. 

\subsection{Algorithm List} \label{app_list}
The following algorithms were experimented in this paper. We constructed training data for the ETRM using these algorithms, and we also created a dataset for evaluation. 
The Clustering Coefficient and Random Walk algorithms were used only in model evaluation. 

\subsubsection{All Vertices In/Out-degree (AID, AOD)}
This algorithm calculates the in/out-degree of all vertices. 
All workers calculate the local degree, and the master-worker calculates the final result by aggregating the local results.

\subsubsection{PageRank (PR)}
This computes the PageRank\cite{page1999pagerank} score of each vertex. 
PageRank score is used to find famous web pages. 
Google's search engine adopted PageRank score. 
The PageRank score is determined by the following equation. $d$ is a damping factor, which is generally 0.85, and the scores of vertices with edges toward v are divided by each out-degree and summed. 
The number of iteration in this paper is 10. 

\begin{align}
    {PR}_{t}(v) = (1-d) + d * \sum_{u\in {N}_{in}(v)}^{}\frac{{PR}_{t-1}(u)}{\left|{N}_{out}(u) \right|}
\end{align}
\subsubsection{Greedy Graph Coloring (GC)}
Graph coloring is an algorithm that changes the color of each vertex so that all neighboring vertices have a different color. 
This algorithm is used in scheduling. 
Minimal graph coloring, which paints the entire vertices with minimal color, is replaced by greedy graph coloring because minimal graph coloring is an NP-complete problem\cite{karp1972reducibility}. We implemented this algorithm based on \cite{10.1007/11823285_61} using the greedy method. 

\subsubsection{All Pair Common Neighbor (APCN)}
This task finds all common neighbor vertices of all pairs of all vertices. 
Two vertices sharing many neighbors can be interpreted that they share similar characteristics in a graph. 
A list of common neighbors can be used in recommendation systems. 

\subsubsection{Triangle Count (TC)}
Regardless of the edge direction, three vertices connected to each other mean there is a triangle. 
This algorithm counts all triangles in the graph data.
The triangle count algorithm is used in social network analysis, especially in detecting communities. 

\subsubsection{All Local Clustering Coefficient (CC)}
In this task, the clustering coefficient of each vertex is computed. This indicates how closely the vertex is bound to neighbor vertices. The higher the coefficient, the more concentrated the vertices. The formula of the local clustering coefficient is as follows. $k_{i}$ represents the number of neighbor vertices of vertex $i$, and $N(i)$ is the neighbor vertices of vertex $i$.
\begin{align}
CC_{i} = \frac{\{e_{jk}:v_j,v_k\in N(i), e_{jk}\in E\}}{k_i(k_i-1)}
\end{align}

\subsubsection{Random Walk (RW)}

Random Walk is an operation of creating random moving samples along the edges of a graph. These samples are used in graph learning. 
To make a sample, the following process is performed. Starting from one source vertex, a random vertex is selected from neighbor vertices connected by out-edges of source vertex. Then, after changing the pivot to the selected vertex, the same process is repeated 10 times for each vertex to form a sample.



\subsection{Evaluation Metrics}
When the graph data and algorithm are configured, each partitioning strategy has its own execution time. 
Let the fastest, the slowest, and average execution time among the execution times of partitioning strategies be $T_{best}$, $T_{worst}$, and $T_{avg}$ respectively. 
The execution time of the partitioning strategy selected by ETRM is $T_{sel}$. 
The followings are our evaluation metrics:

First, $Score_{best}$ compares the fastest partitioning strategy with the selected one. This value has a value from 0 to 1. 
\begin{align}
Score_{best}= \frac{T_{best}}{T_{sel}}
\end{align}
Second, $Score_{worst}$ is a comparison of the slowest partitioning strategy and the selected one. This value has a value greater than 1. 
\begin{align}
Score_{worst} = \frac{T_{worst}}{T_{sel}}
\end{align}
Finally, we compare with the average execution time. This value is always greater than zero. 
\begin{align}
Score_{avg} = \frac{T_{avg}}{T_{sel}}
\end{align}
The higher the score, the better the selected partitioning strategy. 

Our test set contains 96 tasks and it is divided into the following items.
\begin{itemize}
    \item Test set A has 8 tasks and consists of graph data and algorithms that were never used to create the augmented training dataset. 
    \item Test set B has 24 tasks and consists of graph data that was never used to create the augmented training dataset and the algorithms used to create the augmented training dataset. 
    \item Test set C has 16 tasks and consists of algorithms that were never used to create the augmented training dataset and the graph data used to create the augmented training dataset. 
    \item Test set D has 48 tasks and consists of graph data and algorithms used to create the augmented training dataset but not included in this dataset. 
\end{itemize}

\subsection{Evaluation of Selected Strategy}
\input{05_-01chosen_strategy}
\subsection{Importance of Features}
We calculated the Gain importance and Split importance of ETRM's input features. 
Each feature has an average Gain value in ensemble trees. 
Gain importance of a feature is the ratio of the average Gain value of the feature to the sum of average Gain values of all features. 
The split importance of a feature is the sum of the number of times split with the feature. 
Each Table \ref{tab:data_features} and Table \ref{tab:algo_features} contains Gain importance and Split importance. 
The top 4 of the Gain importance ranking are \texttt{Out-degree}, \texttt{The number of Edge}, \texttt{The number of Vertex} and \texttt{In-degree} in order.
All features are Data features. 
The top 4 of the Split importance ranking are \texttt{SUBTRACT}, \texttt{VERTEX\_VALUE\_WRITE}, \texttt{GET\_OUT\_VERTEX\_FROM} and \texttt{OTHERS\_VALUE\_WRITE} in order.
These features are included in Algorithm features. 
We interpreted that the top 4 of the Gain importance ranking help classify input records in the first few levels, and the top 4 of the Split importance ranking help precisely predict the execution time in the last few levels. 
Therefore, we supposed that both data features and algorithm features are important in predicting the execution time. 


\subsection{Benefit-Cost Ratio}
\input{table/benefit_cost_ratio}
We calculated the benefit-cost ratio (BC ratio). 
Benefit means the difference between the performance of the selected strategy and that of the worst strategy in a task. 
Cost is the sum of the elapsed time spent to select a partitioning strategy. 
It is also the same with the sum of time spent to extract data features, time spent to extract algorithm features and time ETRM spent to select the best strategy. 
Cost does not include time spent to write pseudocode of algorithm or time spent to train the ETRM. 
Table \ref{tab:benefit_cost_ratio} show BC ratio of all test tasks. 
In a cell, the number on the top means benefit and the number at the bottom means BC ratio. 

Extracting data features means calculating each feature in Table \ref{tab:data_features}. 
Extracting time varies with the size of graph data. 
The average and variance of the time spent to analyze algorithm code are 0.7 and 0.002539 seconds. 
It took 0.0304 seconds for ETRM to predict the execution time and select a strategy. 

We observed that ETRM has a high BC ratio for algorithms that takes a long execution time in Table \ref{tab:benefit_cost_ratio}. 
For example, all BC ratios are bigger than 1 for \textbf{PR} algorithm but smaller than 1 for \textbf{AID} and \textbf{AOD} algorithms. 
We expect that if we make more samples for \textbf{RW} algorithm, BC ratios will be bigger than 1. 
Some BC ratios are bigger than 1 for \textbf{GC}, \textbf{APCN}, \textbf{TC} and \textbf{CC} algorithms. 
Especially, the benefit is about 2400 seconds and the BC ratio is about 403 for \textit{stanford} graph data and \textbf{APCN} algorithm. 
We suppose that our approach will be more helpful for huge graph data and algorithms that take a long execution time in the real-world. 

\subsection{Summary}
Our experimental study verifies that our approach can select highly ranked, time-efficient strategies. 
We set our evaluation metrics to compare the performance of strategies. 
Selected strategies have 0.95 $\times$ the performance of the best strategies. 
We calculated feature importance to find features that played an important role in the prediction of execution time. 
In addition, we calculated the BC ratio to compare the time saved by selecting a highly ranked strategy and the time spent by ETRM. 

%% file: table/dataset2.tex
\begin{table}[t]
\centering
\caption{Graph data used in experiments}
\label{tab:dataset_table2}
\resizebox{\columnwidth}{!}{%
\begin{tabular}{cccc}
\toprule
Graph Data Name & Vertices & Edges & Direction \\
\midrule
Ego-Facebook (\textit{facebook}) & 4,039 & 88,234 & Undirected \\
Wiki-Vote (\textit{wiki}) & 7,115 & 103,689 & Directed \\
Epinions (\textit{epinions}) & 75,879 & 508,837 & Directed \\
Amazon0312 (\textit{amazon-1}) & 400,727 & 3,200,440 & Directed \\
Slashdot (\textit{slashdot}) & 77,350 & 516,575 & Directed \\
Amazon (\textit{amazon-2}) & 334,863 & 925,872 & Undirected \\
DBLP (\textit{dblp}) & 317,080 & 1,049,866 & Undirected \\
RoadNet-CA (\textit{road-ca}) & 1,965,206 & 2,766,607 & Undirected \\
Gemsec-Deezer-RO (\textit{gd-ro}) & 41,773 & 125,826 & Undirected \\
Gemsec-Deezer-HU (\textit{gd-hu}) & 47,538 & 222,887 & Undirected \\
Gemsec-Deezer-HR (\textit{gd-hr}) & 54,573 & 498,202 & Undirected \\
Web-Stanford (\textit{stanford}) & 281,903 & 2,312,497 & Directed \\
\bottomrule
\end{tabular}%
}
\end{table}


%% file: 05_-01chosen_strategy.tex
This section shows the evaluation of the selected strategy. 
We evaluated the actual ranks of selected strategies. 
In addition, we compared the performance scores of the selected strategies and that of other strategies. 
All evaluations were made on the original 528 real records, not on the augmented training dataset. 

\subsubsection{Rank Evaluation of Selected strategy}
\input{figure/rank_evaluation}

We represent the cumulative ratio of selected strategies' actual rank in Figure \ref{fig:rank_eval}. Because 11 partitioning strategies are used, the cumulative ratio of rank 11 is always 1.0. 
The larger the area of cumulative ratio figure is, the better the selected strategy is.  

Figure \ref{fig:rank_eval1} shows the overall result of test set. 
The ETRM selected the best partitioning strategy in 52\% of the test set and the strategy within rank 4 in 92\% of the test set. 
The areas of figures are large in order of C, D, B, A. It means that ETRM selected better strategies in the same order. 
We expected that ETRM would select better partitioning strategies in test set D. 
However, ETRM showed better selection in test set C. 
ETRM could select the best partitioning strategy only in 33\% of test set B. 
We found that selecting the best strategy in tasks consisting of new graph data is more challenging than selecting the best strategy in tasks consisting of new algorithms. 
When an augmented training dataset is created, synthetic algorithms are aggregated with original algorithms. We interpreted that this is the reason why ETRM could select the best strategy in tasks consisting of new algorithms. 
On the other hand, even the number of augmented training dataset records is about 0.43 million, the number of unique features set of the graph data is only 8. This made ETRM harder to select the best strategy in tasks consisting of new graph data. Therefore, the result of test set B is worse than that of test set C and D. 
The ETRM could not select the best strategy in the test set A, but selected the strategy within rank 4 in 75\% of the test set A. 
The reason is that the test set A is farther away from the augmented training dataset than B, C, and D. 

\subsubsection{Performance Evaluation of Selected strategy}
\input{figure/all_detail_result}
We evaluated the performance of the selected strategy. 
First, we show the performance scores of the selected strategies using box plot. 
Second, Table \ref{table:score_summary} shows the summarized scores. 
Finally, Figure \ref{figure:histogram} has two components. One is comparisons of the performance of the randomly chosen strategy and that of the best strategy. The other one is comparisons of the performance of the selected strategy and the best strategy. 

In Figure \ref{figure:all_box_result}, we depict $Score_{best}$, $Score_{worst}$, and $Score_{avg}$ of the test set. 
We used box plot. In Figure \ref{figure:all_box_result}, each boxes' five points mean minimum, first quartile, median, third quartile and maximum from bottom to top, except outliers. Black triangles are mean values. The vertical red line means whether the graph data or algorithms are used to create the augmented training dataset. 
New graph data and algorithms are in right side of the red lines. 

Figure \ref{figure:data_box_result} is sorted with graph data and Figure \ref{figure:algo_box_result} is sorted with algorithms. 
We observed that figure's result is same with the result of the rank evaluation. 
In $Score_{best}$ subplot of Figure \ref{figure:data_box_result}, mean values of new graph data were lower than that of used graph data except \textit{gd-hr}. 
Whereas, we could not observe drops in mean values when selecting the best strategy for new algorithms. 
$Score_{best}$ is much higher than 
About graph data \textit{amazon-2} and \textit{dblp}, 
Values of $Score_{best}$, $Score_{worst}$ and $Score_{avg}$ in \textit{amazon-2} are closer to 1 than values in other graph data. 
This means the performance of strategies in \textit{amazon-2} have lower variance than in case of other graph data. 
The scores of \textbf{GC} algorithm show a similar pattern. 

\input{table/score_mean_table}
Table \ref{table:score_summary} shows mean values of all scores based on the test set. 
ETRM could select strategies that have 0.95 X performance of the best strategies and outperform twice than the performance of the worst strategies. 
Selected strategies have 0.95 $\times$ the performance of the best strategies, 2 $\times$ the performance of the worst strategies, and 1.46 $\times$ the average performance. 

\input{figure/histogram}
We randomly picked partitioning strategies in the comparison group because manually identifying the best partitioning strategy is infeasible. 
In detail, we randomly picked a partitioning strategy five times for each task and got the mean performance. 
We compared the $Score_{best}$ of randomly chosen strategies and selected strategies by ETRM in Figure \ref{figure:histogram}. 
The strategy that matches the performance of the best strategy within 5\% difference was picked only once in comparison group.  
ETRM selected strategies that match the performance of the best strategy within 5\% difference in 63 tasks. 
Randomly picked strategies have 0.69 $\times$ the performance of the best strategies on average in the whole test set. 
In contrast, ETRM selected strategies that have 0.946 $\times$ the performance of the best strategies on average in the whole test set. 

%% file: figure/rank_evaluation.tex
\begin{figure}[t]
    \centering
    \begin{subfigure}[b]{0.8\columnwidth}
        \centering
        \includegraphics[width=0.8\textwidth]{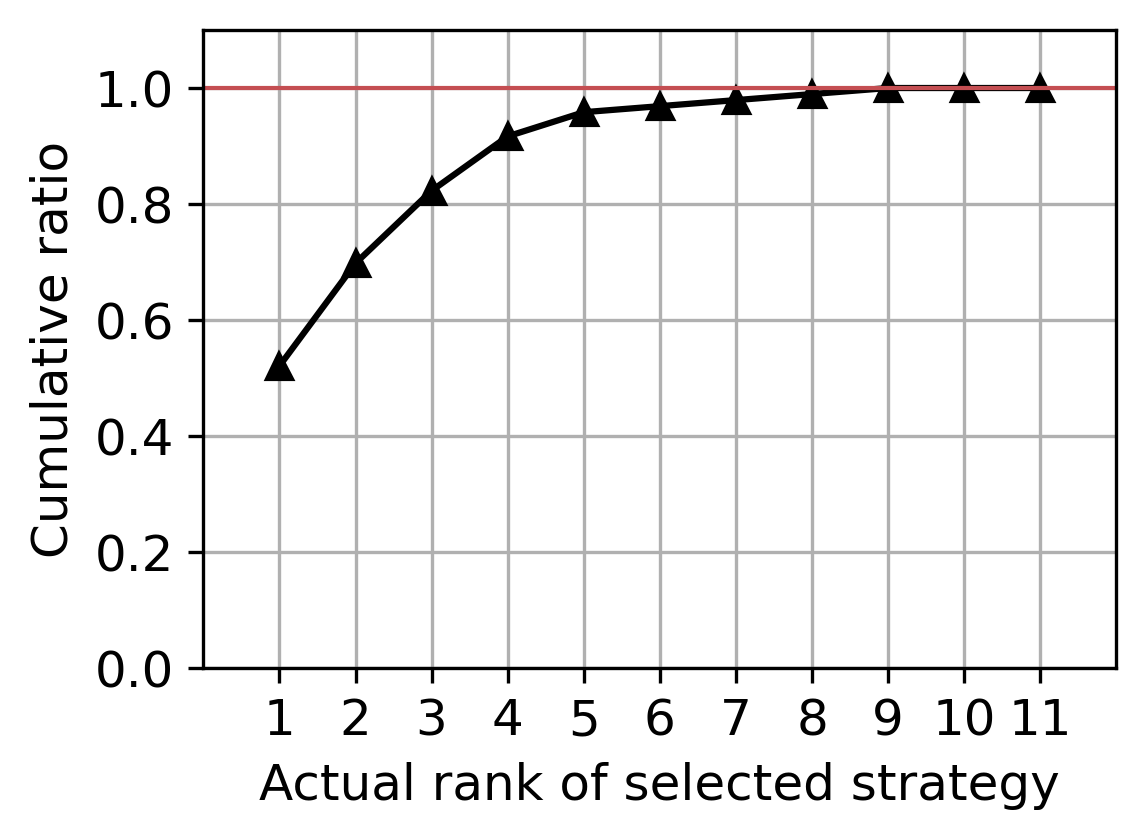}
        \caption{Overall test set}    
        \label{fig:rank_eval1}
    \end{subfigure}
    \newline
    \begin{subfigure}[b]{0.48\columnwidth}  
        \centering 
        \includegraphics[width=\columnwidth]{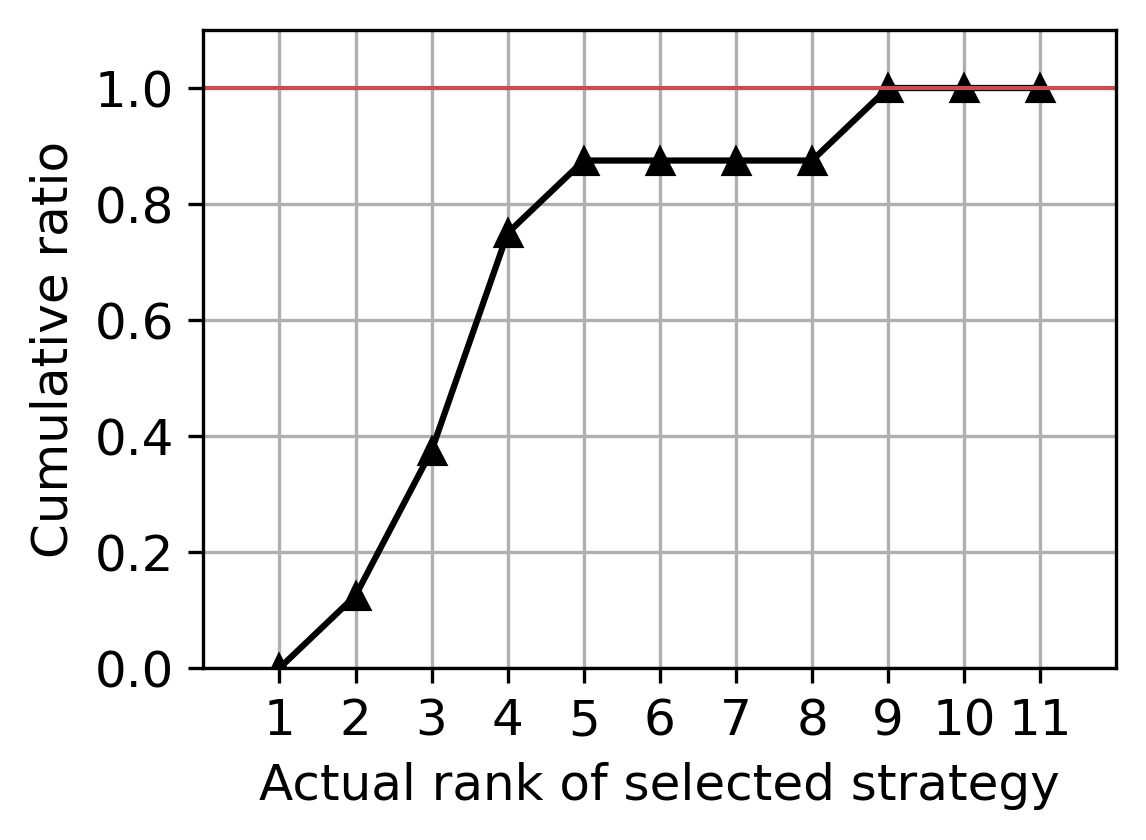}
        \caption[]%
        {Test set A}    
        \label{fig:rank_eval2}
    \end{subfigure}
    \hfill
    \begin{subfigure}[b]{0.48\columnwidth}   
        \centering 
        \includegraphics[width=\columnwidth]{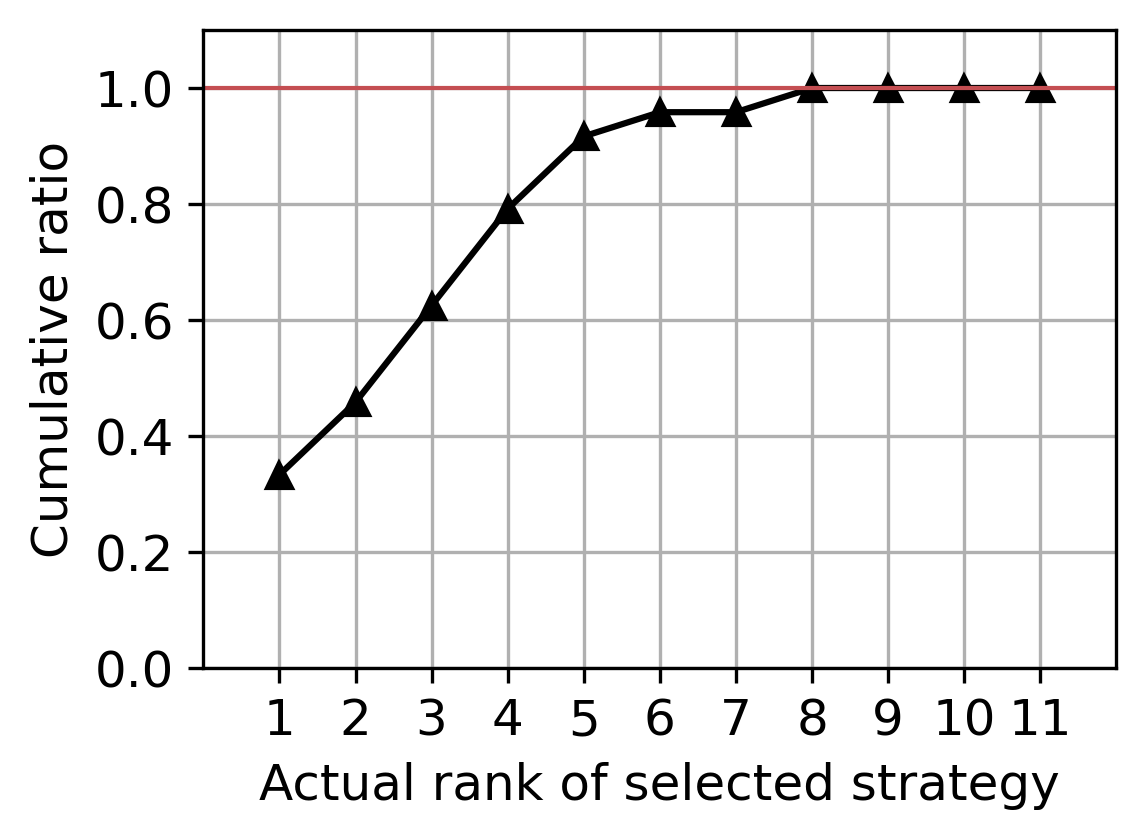}
        \caption[]%
        {Test set B}    
        \label{fig:rank_eval3}
    \end{subfigure}
    \newline

    \begin{subfigure}[b]{0.48\columnwidth}
        \centering
        \includegraphics[width=\columnwidth]{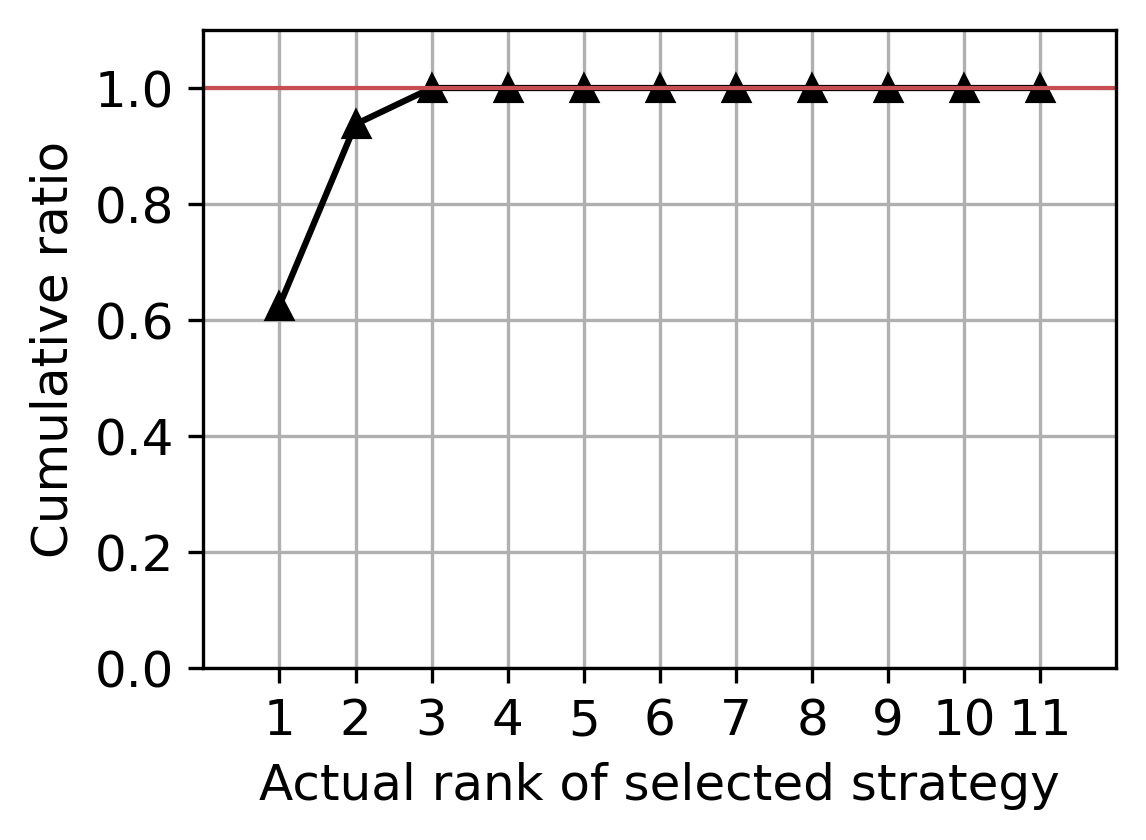}
        \caption[Network2]%
        {Test set C}    
        \label{fig:rank_eval4}
    \end{subfigure}
    \hfill
    \begin{subfigure}[b]{0.48\columnwidth}  
        \centering 
        \includegraphics[width=\columnwidth]{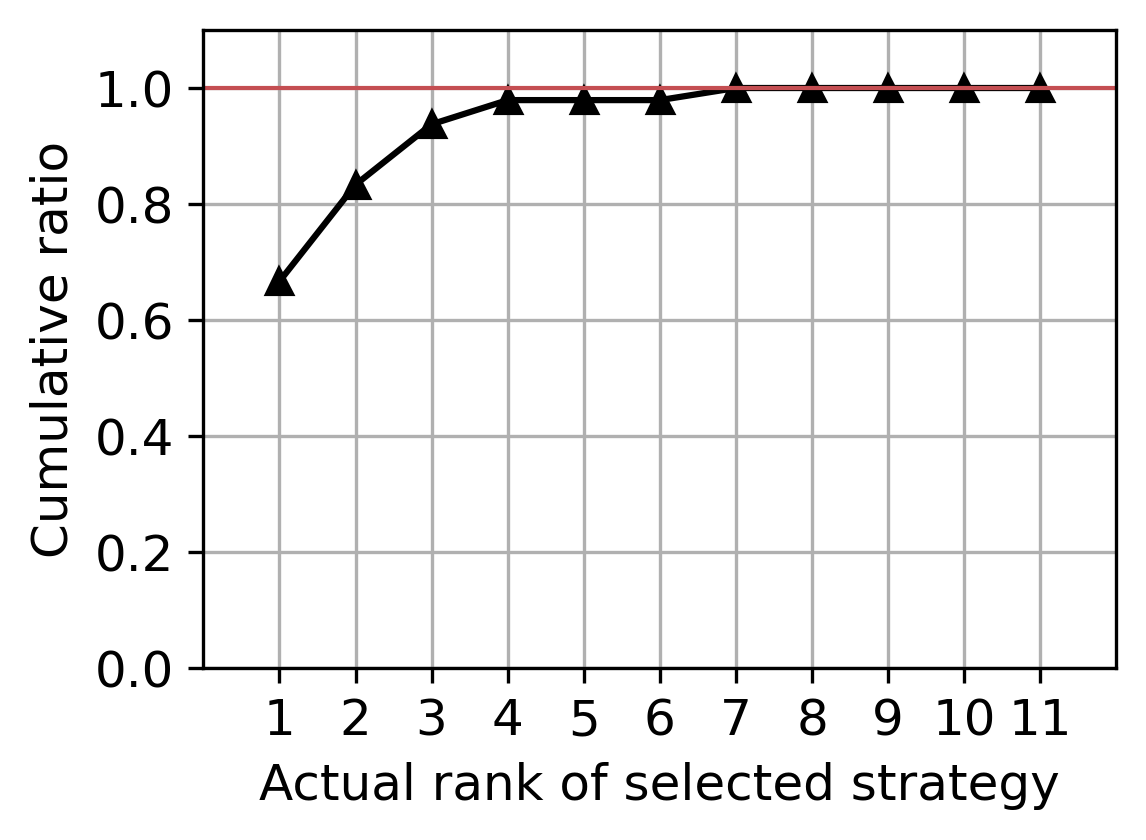}
        \caption[]%
        {Test set D}
        \label{fig:rank_eval5}
    \end{subfigure}

    \caption[Performance of Partitioning Strategies]
    {Cumulative ratio of selected strategies' actual rank}
    
    \label{fig:rank_eval}
\end{figure}

%% file: figure/all_detail_result.tex
\begin{figure*}[!ht]
\centering
\includegraphics[width=0.97\textwidth,height=0.80\textheight,keepaspectratio]{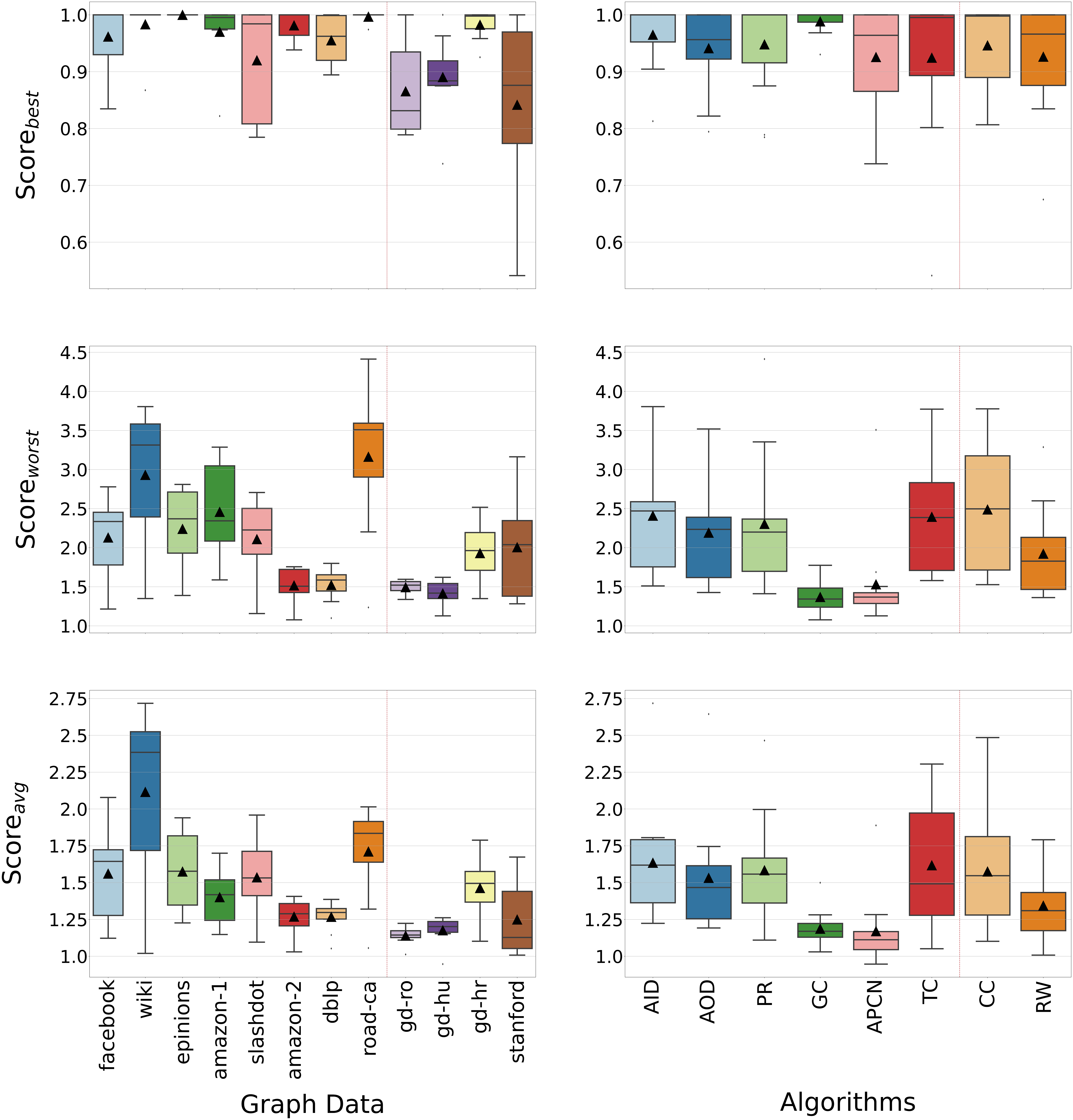}
\begin{minipage}[t]{.5\linewidth}
\centering
\subcaption{Evaluation score on graph data}\label{figure:data_box_result}
\end{minipage}%
\begin{minipage}[t]{.5\linewidth}
\centering
\subcaption{Evaluation score on algorithms}\label{figure:algo_box_result}
\end{minipage}
\caption[Man a woman]{Evaluation score \label{figure:all_box_result}}
\end{figure*}

%% file: table/score_mean_table.tex
\begin{table}[t]
\centering
\caption{Score summary}
\label{table:score_summary}
\begin{tabular}{cccc}
\toprule
\textbf{ }                & $Score_{best}$   & $Score_{worst}$ & $Score_{avg}$  \\ 
\midrule
All cases  & 0.9458 & 2.0770 & 1.4558  \\ 
\hdashline
Test set A & 0.8571 & 1.7905 & 1.2115  \\ 
Test set B & 0.9078 & 1.6857 & 1.2730  \\ 
Test set C & 0.9760 & 2.4146 & 1.5841  \\ 
Test set D & 0.9695 & 2.2079 & 1.5453  \\ 
\bottomrule
\end{tabular}
\end{table}

%% file: figure/histogram.tex
\begin{figure}[t]
\begin{center}
  \includegraphics[width=0.73\columnwidth]{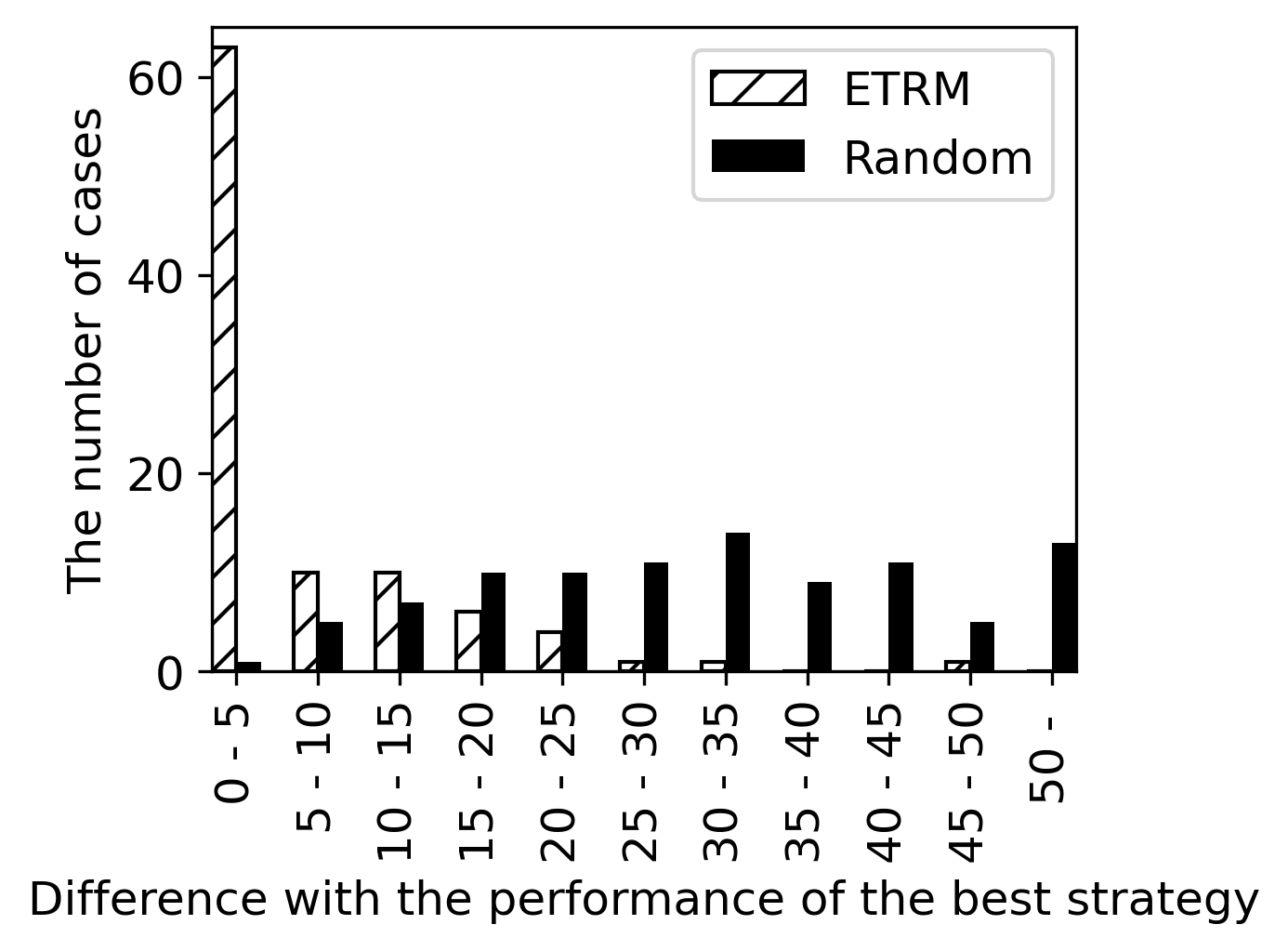}
  \caption{Case count histogram within the difference range from $T_{best}$}
  \label{figure:histogram}
\end{center}
\end{figure}

%% file: table/benefit_cost_ratio.tex
\begin{table}[t]
\centering
\caption{Benefit and Benefit-Cost ratio of the Test set}
\label{tab:benefit_cost_ratio}
\resizebox{\columnwidth}{!}{%

\begin{tabular}{ccccccccc}
 &  &  &  &  & \multicolumn{4}{c}{Benefit (sec), Benefit-cost Ratio (ratio)} \\
 \toprule
 
 & \textbf{AID} & \textbf{AOD} & \textbf{PR} & \textbf{GC} & \textbf{APCN} & \textbf{TC} & \textbf{CC} & \textbf{RW} \\
 \midrule
 
\multirow{2}{*}{\textit{facebook}} & 0.0930 & 0.1480 & 1.5547 & 0.7072 & 0.7463 & 0.5830 & 0.5537 & 0.1767 \\
 & 0.1049 & 0.1671 & \textbf{1.7548} & 0.7982 & 0.8424 & 0.6580 & 0.6249 & 0.1995 \\
 \hdashline
\multirow{2}{*}{\textit{wiki}} & 0.1164 & 0.1922 & 1.8625 & 0.5631 & 1.6831 & 0.7464 & 0.7605 & 0.2288 \\
 & 0.1401 & 0.2315 & \textbf{2.2428} & 0.6781 & \textbf{2.0268} & 0.8989 & 0.9158 & 0.2755 \\
 \hdashline
\multirow{2}{*}{\textit{epinions}} & 0.5640 & 0.8775 & 7.0589 & 2.9990 & 11.9848 & 3.5934 & 3.2175 & 0.9750 \\
 & 0.3536 & 0.5502 & \textbf{4.4260} & \textbf{1.8804} & \textbf{7.5146} & \textbf{2.2531} & \textbf{2.0174} & 0.6114 \\
 \hdashline
\multirow{2}{*}{\textit{amazon-1}} & 5.6319 & 7.2783 & 74.3059 & 29.5124 & 24.0993 & 28.3351 & 29.8422 & 15.8097 \\
 & 0.4305 & 0.5564 & \textbf{5.6804} & \textbf{2.2561} & \textbf{1.8423} & \textbf{2.1661} & \textbf{2.2813} & \textbf{1.2086} \\
 \hdashline
\multirow{2}{*}{\textit{slashdot}} & 0.8842 & 1.2212 & 13.1211 & 4.1318 & 11.0396 & 5.7428 & 5.1091 & 1.6032 \\
 & 0.3871 & 0.5347 & \textbf{5.7453} & \textbf{1.8092} & \textbf{4.8338} & \textbf{2.5146} & \textbf{2.2371} & 0.7020 \\
 \hdashline
\multirow{2}{*}{\textit{amazon-2}} & 0.6525 & 0.9687 & 14.9986 & 2.7370 & 1.7853 & 2.8117 & 2.8536 & 1.3650 \\
 & 0.1259 & 0.1870 & \textbf{2.8949} & 0.5283 & 0.3446 & 0.5427 & 0.5508 & 0.2635 \\
 \hdashline
\multirow{2}{*}{\textit{dblp}} & 1.0324 & 1.3911 & 13.7510 & 3.4462 & 1.5099 & 2.7971 & 2.8312 & 1.6562 \\
 & 0.1917 & 0.2583 & \textbf{2.5535} & 0.6399 & 0.2804 & 0.5194 & 0.5257 & 0.3076 \\
 \hdashline
\multirow{2}{*}{\textit{road-ca}} & 14.0050 & 19.8851 & 303.7410 & 48.5656 & 27.7334 & 45.8575 & 45.3406 & 23.4419 \\
 & 0.3576 & 0.5078 & \textbf{7.7558} & \textbf{1.2401} & 0.7082 & \textbf{1.1709} & \textbf{1.1577} & 0.5986 \\
 \hdashline
\multirow{2}{*}{\textit{gd-ro}} & 0.1194 & 0.1739 & 1.9616 & 1.3282 & 0.1978 & 0.4016 & 0.3877 & 0.3181 \\
 & 0.0936 & 0.1364 & \textbf{1.5385} & \textbf{1.0417} & 0.1551 & 0.3150 & 0.3041 & 0.2495 \\
 \hdashline
\multirow{2}{*}{\textit{gd-hu}} & 0.2121 & 0.2471 & 2.4513 & 1.1798 & 0.1379 & 0.6543 & 0.5868 & 0.3237 \\
 & 0.1298 & 0.1512 & \textbf{1.5000} & 0.7219 & 0.0844 & 0.4004 & 0.3591 & 0.1981 \\
 \hdashline
\multirow{2}{*}{\textit{gd-hr}} & 0.7474 & 1.0651 & 9.9416 & 2.2850 & 1.2029 & 2.0912 & 2.1406 & 1.0360 \\
 & 0.3690 & 0.5259 & \textbf{4.9085} & \textbf{1.1282} & 0.5939 & \textbf{1.0325} & \textbf{1.0569} & 0.5115 \\
 \hdashline
\multirow{2}{*}{\textit{stanford}} & 2.4827 & 3.0283 & 41.1078 & 9.2639 & 2419.3177 & 80.9914 & 96.5416 & 2.5428 \\
 & 0.4139 & 0.5048 & \textbf{6.8529} & \textbf{1.5443} & \textbf{403.3138} & \textbf{13.5017} & \textbf{16.0940} & 0.4239 \\
 \bottomrule
\end{tabular}%
}
\end{table}

%% file: 02_related_work.tex
\section{Related Work}


\subsection{Distributed Graph Computation Engine}
Recent studies have introduced a distributed graph processing system to overcome the limitation of a single node and maximize parallelism. 
Giraph\cite{10.14778/2777598.2777604}, GraphX\cite{10.5555/2685048.2685096, 10.1145/2484425.2484427}, and Gelly\cite{10.1145/3167132.3167147} provided libraries that can perform graph algorithms in Hadoop, Spark, and Flink, which are general-purpose frameworks that allow users to perform distributed programming easily. 
These distributed framework-based graph systems express diverse iterative graph algorithms using simple programming abstraction and aim for linear-scalable execution rather than optimizing single iteration. 
Based on the same philosophy as the above system, Pregel\cite{10.1145/1807167.1807184} presented a think like a vertex model that defines what behavior should be performed from the vertex perspective for each iteration, and GraphLab\cite{10.5555/3023549.3023589} and Cyclops\cite{10.1145/2600212.2600233} also used this vertex-centric model. 
However, in many real-world graph data has skewness property that a few vertices are connected to a large number of edges, whereas most vertices are connected to a small number of edges. 
PowerGraph\cite{10.5555/2387880.2387883} pointed out that the vertex-centric model has a limitation to balancing computation in the distributed processing of these power-law graphs, and proposed a Gather-Apply-Scatter (GAS) model that parallelizes computation based on the edge. 
In addition, PowerLyra\cite{chen2019powerlyra} introduced a hybrid approach that differentiates computation and partitioning strategies according to the degree of the vertex for processing skewed graphs. 
\subsection{Partition Strategies}
Data partitioning has a major effect on data locality, load balancing, and replication factor in a distributed environment. 
One way to partition a graph is an edge-cut method that assigns vertices to each node and spans edges that connect vertices belonging to different nodes. 
Another alternative is a vertex-cut method in which the necessary vertices are replicated and all edges are distributed to each node. 
These two methods have contradictory advantages and disadvantages. 
In edge-cut, high degree vertex causes workload imbalance and vertex-cut increases replication factor of spanning vertex\cite{chen2019powerlyra}. 
Previous studies have suggested variations to compensate for the shortcomings of each edge-cut and vertex-cut. 
METIS\cite{karypis1998fast} proposed a multi-level partitioning algorithm to reduce the number of ghost vertices in edge-cut, and greedy heuristic algorithm\cite{10.5555/2387880.2387883} reduced the replication factor of vertex-cut. 
PowerLyra\cite{chen2019powerlyra} proposed hybrid approaches that use several partitioning algorithms together.

\subsection{Application Driven Re-Partitioning}
There has also been a study to improve performance through re-partitioning. The goal of \cite{10.1145/3318464.3389745} is to re-partition graph data to suit a specific application. \cite{10.1145/3318464.3389745} uses 
the application's execution time record to make the cost function as close to this record as possible. First, after performing the first partitioning process, the cost function model predicts each worker's computation cost and communication cost. After that, some partial data of workers with high predicted costs are transferred to workers with lower predicted costs. This process is a re-partitioning process. All workers are cost balanced, so the task gets done faster. 
Model of \cite{10.1145/3318464.3389745} can only re-partition graph data for the already trained algorithm. 
To re-partition graph data for a new algorithm, a new model must be trained. 

\subsection{Partitioning Advisor for Databases}
There have been several studies \cite{10.1145/1066157.1066292, 10.1145/564691.564757, 10.1145/1989323.1989444, 10.1145/3318464.3389704} that configure the physical setting of a database automatically, including partitioning data. 
In these studies, each cost model is used to evaluate query performance and change settings to improve performance. 
Especially, \cite{10.1145/3318464.3389704} train their model with the log of query execution using reinforcement learning and advise the partition to be appropriate for the query workload. 
The difference from our research is that the studies presented above are focused on distributed databases, and our studies are focused on distributed graph computation. 
It is necessary to consider the connected vertices and edges. 
The relational database system has a query optimizer and formalized syntax SQL. 
These functions can help to compute the cost of a query. However, distributed graph processing engine does not have these functions and has to compute execution cost in another way. 

%% file: 06_conclusion.tex
\section{Conclusion}
We propose a method to select the most appropriate partitioning strategy for a given graph in a specific algorithm by estimating its execution time. 
Selected strategies have 0.95 $\times$ the performance of the best strategies. 
Benefit-Cost ratio in \textit{Web-stanford} data and \textbf{APCN} algorithm is about 403. 
It shows that our work is much helpful for huge graph data and algorithms that take a long execution time in practice. 
For our future work, 1) We will find other effective features of graph data and algorithms. 2) We will test our approach in other distributed graph computation models. 3) We are planning to design a distributed system that exploits the result of our research effectively.

%% file: 07_acknowledgement.tex
\section*{ACKNOWLEDGEMENT}
 This work was supported by the National Research Foundation of Korea (NRF) grant funded by the Korea government (the  Ministry of Science and ICT, MSIT) (2016M3C4A7952630, Genome Scale Protein Structure Modeling). 